# Risk Attitudes and Human Mobility during the COVID-19 Pandemic

Ho Fai Chan[1,2*], Ahmed Skali[3], David Savage[2,4], David Stadelmann[2,5,6,7], Benno Torgler[1,2,7]

[1]School of Economics and Finance, Queensland University of Technology, Australia
[2]Centre for Behavioral Economics, Society and Technology (BEST), Australia
[3]Department of Economics, Deakin University, Australia
[4]Newcastle Business School, University of Newcastle, Australia
[5]University of Bayreuth, Germany
[6]IREF – Institute for Research in Economic and Fiscal Issues
[7]CREMA—Center for Research in Economics, Management, and the Arts, Switzerland
*e-mail: hofai.chan@qut.edu.au

**Abstract**

Behavioral responses to pandemics are less shaped by actual mortality or hospitalization risks than they are by risk attitudes. We explore human mobility patterns as a measure of behavioral responses during the COVID-19 pandemic. Our results indicate that risk-taking attitude is a critical factor in predicting reduction in human mobility and increase social confinement around the globe. We find that the sharp decline in movement after the WHO (World Health Organization) declared COVID-19 to be a pandemic can be attributed to risk attitudes. Our results suggest that regions with risk-averse attitudes are more likely to adjust their behavioral activity in response to the declaration of a pandemic even before most official government lockdowns. Further understanding of the basis of responses to epidemics, e.g., precautionary behavior, will help improve the containment of the spread of the virus.

*In Thackeray's novel Henry Esmond, for instance, this dread informs the narrative. The heroine, Lady Castlewood, contracts the disease as an adult. Her husband had been a brave soldier in combat, but he was unable to face a malady that he could not fight and that threatened him not only with death, but also with disfigurement. Unwilling to put his pink complexion and his fair hair at risk, Lord Castlewood took to his heels and deserted his household for the duration of the outbreak. But he was not part of a mass exodus, even though Henry Esmond declares that smallpox was "that dreadful scourge of the world" and a "withering blight" and "pestilence" that "would enter a village and destroy half its inhabitants."*

Snowden (2019, p. 101).

The central features of modern global society make us more vulnerable to the challenge of pandemic diseases and their global implications, as viral transmission can trigger large-scale responses (1)[1]. Epidemics such as COVID-19 threaten our social fabric (2), thus it is important to understand such occurrences from a risk behavior perspective. Scholars have emphasized how social and behavioral sciences can offer important insights into how the COVID-19 pandemic may be understood and managed (3). Risk behavior has been predominately analyzed in relation to the HIV/AIDS pandemic (4,5). Studies have also tried to model the effects of risk perception on the spreading of an epidemic (6), and have explored how different levels of awareness may help to prevent an outbreak (7). Other studies have explored the implications of risk attitudes in disasters (8–10) or extreme situations (11).

Risk-taking attitudes and behavior are important elements of human behavior as they determine a range of decision-making strategies (12) and contribute to navigation of the complexity, uncertainty, and dangerous world where risk looms large. For example, research has shown that risk aversion can result in the over-weighting of risk factors and risk-seeking can result in the under-weighting of risk (13–17). Advanced civilizations dating back to the Asipu in Mesopotamia in 3200 B.C. had risk management strategies in place to estimate profits/losses or successes/failures ((18) discussed in (19)). Another early example of risk analysis and risk management is the *Code of Hammurabi*[2] issued in 1950 B.C. (19). Our cognitive apparatus has equipped us evolutionarily to survive our daily activities (20), while enduring and recurring risks in the environment have required evolutionary adaptiveness as a core selective factor of survival (21). The implication is that we must remain safe to guarantee our survival. It is no coincidence that we are all well aware of the proverb "Better safe than sorry".

---

[1] In the Middle Ages, for example, monasteries were vulnerable to plagues. Their status as central hubs meant they acted as nodes in the grain trade; linking villages and settlements together, and attracting a substantial community of people who lived close by. In addition, monasteries served as a place of refuge (1, p. 42).
[2] The *Code of Hammurabi* is Babylonian code of law that is still well-preserved.

Risk entails a complete probabilistic knowledge of something occurring, which allows a decision regarding what action to take. However, not only are we boundedly rational human beings (22) subject to emotions (23) such as fear, but the complexity of the environment and situation, the limited available information on contextual factors of other humans, or dynamic changes may not allow us to have a clear idea about the actual probability we face[3]. In addition, calculating the probability of risk is not the same as actually perceiving it, and humans use less accurate heuristics to make judgments that also include perception of risk. Our biases often disrupt our risk assessments in both positive and negative ways by limiting access to information (searches), limited cognitive understanding (noise), and through our own personal experiences. Thus, subjective perceptions or emotional responses may be triggered by human traits or other factors. For example, we adjudge risk differently based on the physical distance between ourselves and the danger, i.e., we feel safer if the danger is further away, and we are less likely to continuously monitor it over an extended duration (24). This may work relatively well for traditional dangers like fires or floods, but the spread of a pandemic is invisible, and only media reports of those in hospital give any rough clue to its presence. As such, it is likely that we fail to correctly use local transmission (infection) rates as a guide of its proximity or distance to us and the level of threat it poses. Risk as a feeling is less driven by actual probabilities and more by our instinctive and intuitive reaction to danger (20, p. 70). Risk-taking has often been classified as a stable personality trait (25), although situational or contextual factors can also matter (see, e.g., 26–28). An individual's risk type and their perception of risk are highly correlated, such that they interact to exacerbate the underlying risk type. That is: risk seekers are likely to have a worse perception of risk and not only are they willing to accept more gambles, but their estimations of the gambles are underweighted, leading to greater adoption of risk than the individual intended (29). In addition, we humans are also subject to framing biases, reacting differently depending on the way in which information is presented (e.g., positively or negatively, see 30)[4]. This framing can increase or decrease our willingness to take or avoid risk, especially where losses are concerned – the loss of life from contracting the virus is the ultimate loss. Thus, preferences are not set in stone and are open to change, especially after we experience losses; i.e., an individual may be more risk-seeking following losses and risk-averse following gains (10, 31–33).

---

[3] Some frontline professions more exposed to interaction with other people have a higher risk of being infected. Looking historically at plagues, professions such as street vendors, physicians, priests, gravediggers or washerwomen were more seriously at risk of acquiring or transmitting diseases when moving from place to place.

Feelings elicited during a pandemic have an impact on everyday activities (34) and individuals are required to make trade-offs that are affected by their risk behavior. Is it safe to go out shopping, to the park, to use public transportation etc.? What are the chances of getting infected? How do we need to respond? Risk attitudes matter as individuals are aware that going into public places increases the possibility of being infected; if there was to be an infection, this would be subsequently regretted. Risk-averse individuals may respond more to unfamiliar risks that are perceived as uncontrollable (35). During pandemics, states also may become more controlling – historically, social mobility restrictions or regulations are common in pandemics. For example, anti-plague regulations banned funerals, processions, sale of clothing, and gatherings in public assemblies, all of which reduced opportunities for trade, and imposed severe penalties when those rules were not followed. Community bonds might be destroyed if people lose the opportunity to, for example, grieve, pay final respects, or assemble (1). The level of social mobility in our current situation is interesting, as during this phase there is no real treatment or vaccination, which means that citizens need to rely on precautionary behavior. As the reality of the COVID-19 outbreak emerged, we saw that states started to introduce social distancing and isolation measures to deal with the pandemic and the lack of a vaccine.

---

[4] For example, a patient may opt for surgery with a 95% survival rate but not for a surgery with a 5% chance of death.

In this article, we take a look at key social or human mobility factors related to retail and recreation, grocery stores and pharmacies, parks, transit stations, workplaces, and private residences. To measure risk-taking attitude, we use the Global Preference Survey (36,37), which analyzed risk at the country level by combining experimental lottery choice sequences using a staircase method (choice between a fixed lottery in which individual could win x or zero and varying sure payments) and self-assessment based on the willingness to take risks (see Method section for more details). We then extended this data to obtain regional level information. Exploring how risk attitude affects social mobility at the regional level is interesting as risk behavior can be seen as the product of an interplay between individuals, actions of others, and the community or social environment (4). Risk is therefore deeply embedded in specific sociocultural backgrounds (38), with country and geographical differences in risk-taking reported by scholars such as (36) (e.g., higher risk-taking values in Africa and the Middle East while Western European countries are relatively risk-averse). In the context of a pandemic where a population is attempting social isolation or are in lockdown, we see that shopping behaviors change (drop) and large swathes of the workforce have lost their jobs, which means that the entire population has been directly affected by the pandemic if not the virus. It is therefore interesting to explore how citizens' responses to an epidemic are driven by risk attitudes or preferences at the community or regional level.

In particular, we are interested in how individual behavior responses to global announcements – such as the COVID-19 outbreak classification as a pandemic[5] by the WHO – can be shaped by risk attitudes. We suggest that people in risk-taking environments may be less likely to respond and engage in behavioral change which reduces risk. We are also interested in comparing situations with higher or lower opportunity costs in human mobility. The opportunity costs of staying home are defined as the cost incurred by not enjoying the benefit of going out (benefits associated with the best alternative choice). For this, we explore differences between weekdays and the weekend. As many individuals are still working during the week, even while being at home, there is more psychological pressure to be active during the weekend, which increases the opportunity costs of staying at home. Not going out requires more psychological costs to fight against previously formed habits, as it is difficult to abandon the way in which we are accustomed to act. We therefore hypothesize that regions with higher risk attitudes are less likely to follow precautionary strategies when opportunity costs are higher and are therefore are less likely to deviate from their outside activities during the weekend relative to the baseline. Lastly, we also examine whether people adjust their behavior when living in a population with a larger proportion of older people at greater risk of more serious illness from contacting the virus. We expect that regions with a higher share of over 65 individuals would show a greater reduction in mobility. In particular, risk-averse regions may display stronger mobility deviations from their original baseline (stronger reduction).

---

[5] http://www.euro.who.int/en/health-topics/health-emergencies/coronavirus-covid-19/news/news/2020/3/who-announces-covid-19-outbreak-a-pandemic

**Results**

We examined the relationship between the changes in human mobility during the outbreak of Coronavirus disease (COVID-19) and the average risk preferences of individuals in 58 countries (with 776 regions from 33 countries with subnational regions data)[6]. Our main goal is to see if individuals in areas with higher (lower) levels of willingness to take risks are less (more) likely to reduce their exposure to social interactions by going to public places between 15 Feb 2020 and 09 May 2020. The outcome variables measure the daily *changes* (in percentage) in location visits compared to the median value of the same day of the week in the 5-week baseline period, during 3 January and 6 February 2020. To see whether mobility changes are related to risk tolerance, we first regressed the each of the six mobility measures on risk-taking preference, namely, *Retail & Recreation*, *Grocery & Pharmacy*, *Parks*, *Transit Stations*, *Workplaces*, and *Residential*. In each regression, we controlled for whether the day is a weekend, an indicator distinguishing our sample time period by the day when the World Health Organization (WHO) declared the COVID-19 outbreak a pandemic (11 March 2020), the total number of confirmed cases per 1,000 people, number of days since the first confirmed coronavirus related death in the country[7], percentage of population over 65, population density (per squared km of land area), percentage of urban population, average household size, unemployment rate, per capita income (in logs), daily average temperature, and a set of indicators on government responses that covers recommending and requesting closure of school, workplace, public transport, stay at home, cancellation of public events, and restriction on gatherings and internal movement (39). Consequently, our results regarding risk attitudes can be interpreted as independent of government lockdown measures. To this end, we employed a random-effects linear model to estimate the linear effect of risk-preference on mobility and linear interaction effects of risk and other covariates, namely, pandemic declaration, weekend, and the share of population over 65.

---
[6] For most countries, regions are identified as the first-level administrative divisions. For Japan and Great Britain, regions are identified as the second-level administrative divisions.
[7] Days with no deaths (or before a death occurred) coded 0.

As expected, we see an overall reduction in visits to all localities for almost all regions other than residential places, particularly in the earlier weeks in the sample period (see Fig. 1). Interestingly, a large proportion of observations showed an increase in visits to parks, even in the earlier phase. Examining the general relationship between risk attitude and the change in mobility in the entire sample period, we find some evident relationship to two locations. Particularly, risk-taking is positively associated with the change in visitation to places classified as retail and recreation ($\beta$=2.873, s.e.=1.180, CI$_{95\%}$=[0.561;5.185], $P$=0.015) and parks ($\beta$=7.667, s.e.=2.577, CI$_{95\%}$=[2.616;12.718], $P$=0.003), which indicates that in areas with higher average risk-tolerance, an individual is more likely to visit these places (or less likely to reduce their frequency of visits). On the other hand, there is no apparent relationship between risk preference and change in mobility to grocery and pharmacy ($\beta$=-0.481, s.e.=1.060, CI$_{95\%}$=[-2.559;1.597], $P$=0.650), transit stations ($\beta$=1.352, s.e.=1.350, CI$_{95\%}$=[-1.294;3.998], $P$=0.317), workplaces ($\beta$=0.306, s.e.=0.848, CI$_{95\%}$=[-1.355;1.967], $P$=0.718) and residential areas ($\beta$=-0.241, s.e.=0.374, CI$_{95\%}$=[-0.973;0.492], $P$=0.519).

Most control variables report the expected effect on change in human mobility. Specifically, there is a reduction in outings and an increase in staying home as severity increases, such as after the WHO declared coronavirus outbreak a global pandemic, increase in the number of case per population (except for parks and residential, in which the relationship is positive and significant at 10% level and not significant, respectively), and most lockdown measures[8] (see Supplementary Table S1). We also find that, on average, there is a greater reduction in visits to retail and recreational places ($\beta$=-4.386, s.e.=0.132, CI$_{95\%}$=[-4.645;-4.127], $P$<0.001), grocery and pharmacy ($\beta$=-3.969, s.e.=0.167, CI$_{95\%}$=[-4.295;-3.642], $P$<0.001), parks ($\beta$=-4.543, s.e.=0.446, CI$_{95\%}$=[-5.417;-3.669], $P$<0.001), and transit stations ($\beta$=-0.791, s.e.=0.189, CI$_{95\%}$=[-1.162;-0.421], $P$<0.001) on the weekends, in contrast to weekdays, while at the same time a reduction in visits to workplaces ($\beta$=8.277, s.e.=0.215, CI$_{95\%}$=[7.855;8.698], $P$<0.001) and staying home ($\beta$=-3.284, s.e.=0.110, CI$_{95\%}$=[-3.501;-3.068], $P$<0.001) is stronger in weekdays, compared to weekends. We note that while the number of days since the first death in the nation decreases significantly with going to transit stations ($\beta$=-0.387, s.e.=0.138, CI$_{95\%}$=[-0.658;-0.117], $P$=0.005), it had no effect or positive effect on mobility to other localities. Decline in visits to grocery and pharmacy ($\beta$=-0.715, s.e.=0.141, CI$_{95\%}$=[-0.992;-0.438], $P$<0.001), transit stations ($\beta$=-0.304, s.e.=0.166, CI$_{95\%}$=[-0.629;0.022], $P$=0.068), and workplaces ($\beta$=-0.361, s.e.=0.092, CI$_{95\%}$=[-0.542;-0.180], $P$<0.001) is stronger for countries with a higher population density.

---

[8] In almost all cases, mobility is negatively and significantly correlated with the strictest confinement measure (e.g., require closing of all non-essential workplace). The exception being restriction on gatherings, in which the effect is not precisely estimated. Some social isolation recommendations (stay at home and work place) and intermediate restrictions (schools and gatherings) reports the opposite effect on mobility to expectation.

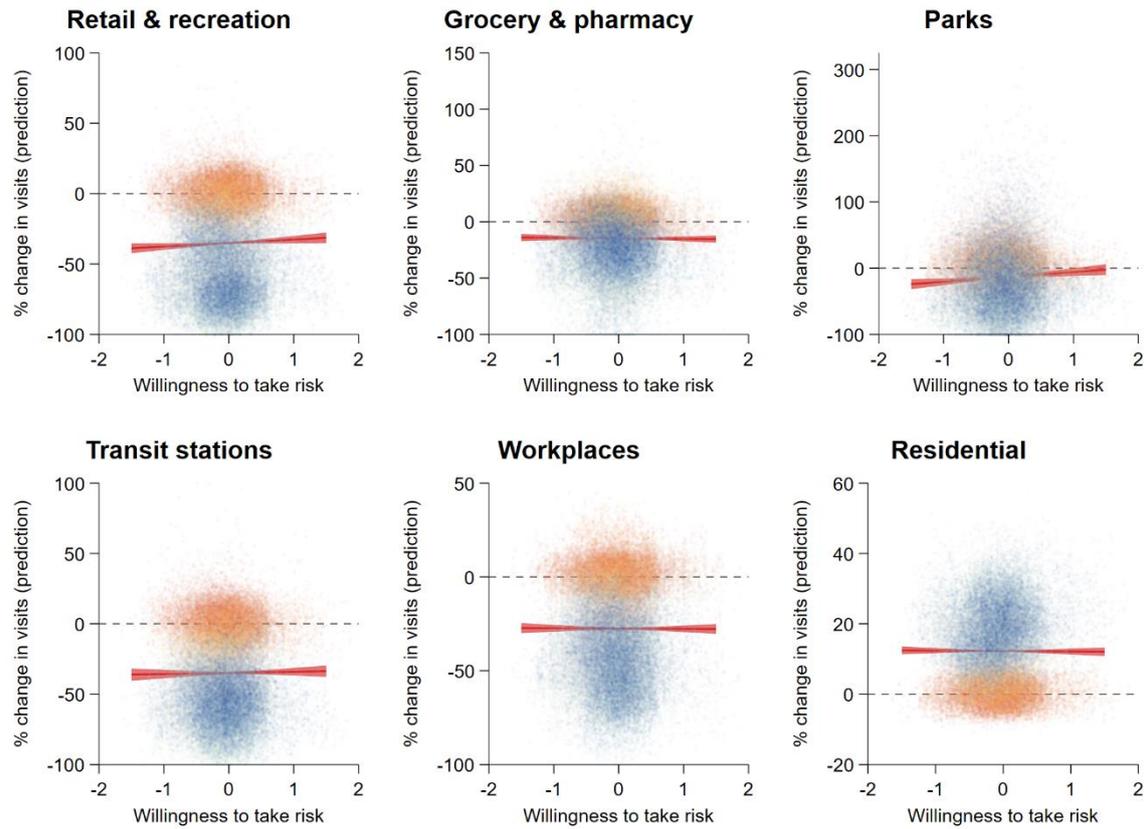

**Fig. 1 | Risk attitude and human mobility during COVID-19.** The six panels show the predicted percentage change in visit to locations classified as retail and recreation ($\beta$=2.873, s.e.=1.180, CI95%=[0.561;5.185], $P$=0.015), grocery and pharmacy ($\beta$=-0.481, s.e.=1.060, CI$_{95\%}$=[-2.559;1.597], $P$=0.650), parks ($\beta$=7.667, s.e.=2.577, CI$_{95\%}$=[2.616;12.718], $P$=0.003), transit stations ($\beta$=1.352, s.e.=1.350, CI$_{95\%}$=[-1.294;3.998], $P$=0.317), workplaces ($\beta$=0.306, s.e.=0.848, CI$_{95\%}$=[-1.355;1.967], $P$=0.718), and residential ($\beta$=-0.241, s.e.=0.374, CI$_{95\%}$=[-0.973;0.492], $P$=0.519), compared to the respective baseline values over average individual risk preference. Estimates of the risk-mobility relation are obtained from random-effects linear regression (Table S1). Markers represent the daily change in visits to the six locations for each region during the entire sample period[9], with different colors showing observations over time (from most blue (first week of the sample period) to yellow (middle of the sample period) to most red (last week of the sample period)).

*Does the pandemic declaration increase the effect of risk-attitude?* We examine the interaction between willingness to take risks and pandemic declaration to assess if the effect of risk-taking on mobility is evident. We find evidence suggesting the declaration is a strong moderator of the risk-mobility effect. It is relevant to note that the declaration of the pandemic precedes lockdown measures of most governments.

---

[9] For visualization purpose, we excluded the Jammu and Kashmir (India) region.

We see that the reduction in outdoor activities (or increase in staying home) can be observed before COVID-19 was declared a pandemic by the WHO, especially for visits to places classified as retail and recreation, transit stations, and workplaces (see Fig. 2). The magnitude of mobility change has indeed increased after the declaration. For example, there is a further 11.3 percentage point (pp) drop in visits to retail and recreation locations ($\beta$=-11.328, s.e.=0.879, CI$_{95\%}$=[-13.051;-9.606], $P$<0.001), 7.3pp drop in going to parks ($\beta$=-7.303, s.e.=1.379, CI$_{95\%}$=[-10.006;-4.600], $P$<0.001), 12pp drop in going to transit stations ($\beta$=-11.998, s.e.=0.833, CI$_{95\%}$=[-13.631;-10.365], $P$<0.001), and 8pp drop in going to workplaces ($\beta$=-8.103, s.e.=0.642, CI$_{95\%}$=[-9.361;-6.846], $P$<0.001), respectively, compared to the period before pandemic declaration (Fig. 2, Table S1). In contrast, we find an average of 3.6pp increase in staying in a residential area ($\beta$=3.602, s.e.=0.285, CI$_{95\%}$=[3.042;4.161], $P$<0.001) after declaration. Interestingly, the pandemic declaration did not have a severe impact on visits to grocery stores and pharmacies ($\beta$=-0.987, s.e.=0.705, CI$_{95\%}$=[-2.368;0.395], $P$=0.162)[10].

We find that, with respect to risk preferences, the changes to visitation patterns (compared to their respective baseline) are relatively greater for areas with lower average willingness to take risk, following the pandemic declaration. Specifically, we find the reduction in visits to grocery and pharmacy, transit stations, and workplaces *prior* to declaration is negatively correlated with willingness to take risk. However, interrogating the interaction terms between risk-taking and pandemic declaration revealed a more interesting behavioral pattern; that is, the additional reduction in out-of-home activities after the declaration is much more dramatic for areas with more less risk-tolerating individuals. We found a statistically significant interaction effect on each of the outcome variables except for residential places (retail and recreation: $\beta$=6.715, s.e.=1.166, $CI_{95\%}$=[4.430;9.001], $P$<0.001; grocery and pharmacy: $\beta$=5.983, s.e.=1.013, $CI_{95\%}$=[3.998;7.968], $P$<0.001; parks: $\beta$=11.910, s.e.=2.449, $CI_{95\%}$=[7.110;16.711], $P$<0.001; transit stations: $\beta$=7.168, s.e.=1.422, $CI_{95\%}$=[4.381;9.954], $P$<0.001; workplaces[11]: $\beta$=4.020, s.e.=0.871, $CI_{95\%}$=[2.313;5.726], $P$<0.001; see Fig. 2). It is also important to note that the pre- and post-declaration change in visitation pattern differences are smaller for higher risk-tolerance areas and vice versa, indicating that areas with higher average risk-taking are less likely to respond to the negative change in environmental status.

---

[10] Nonetheless, the findings in our robustness checks (Table S8) suggest that visits to grocery and pharmacy also decreased significantly after the pandemic declaration, which is also in line with the estimate obtained from Table S1.

---

[11] One should note that in two out of the three robustness checks (Table S8), the interaction effect on workplaces is not precisely estimated.

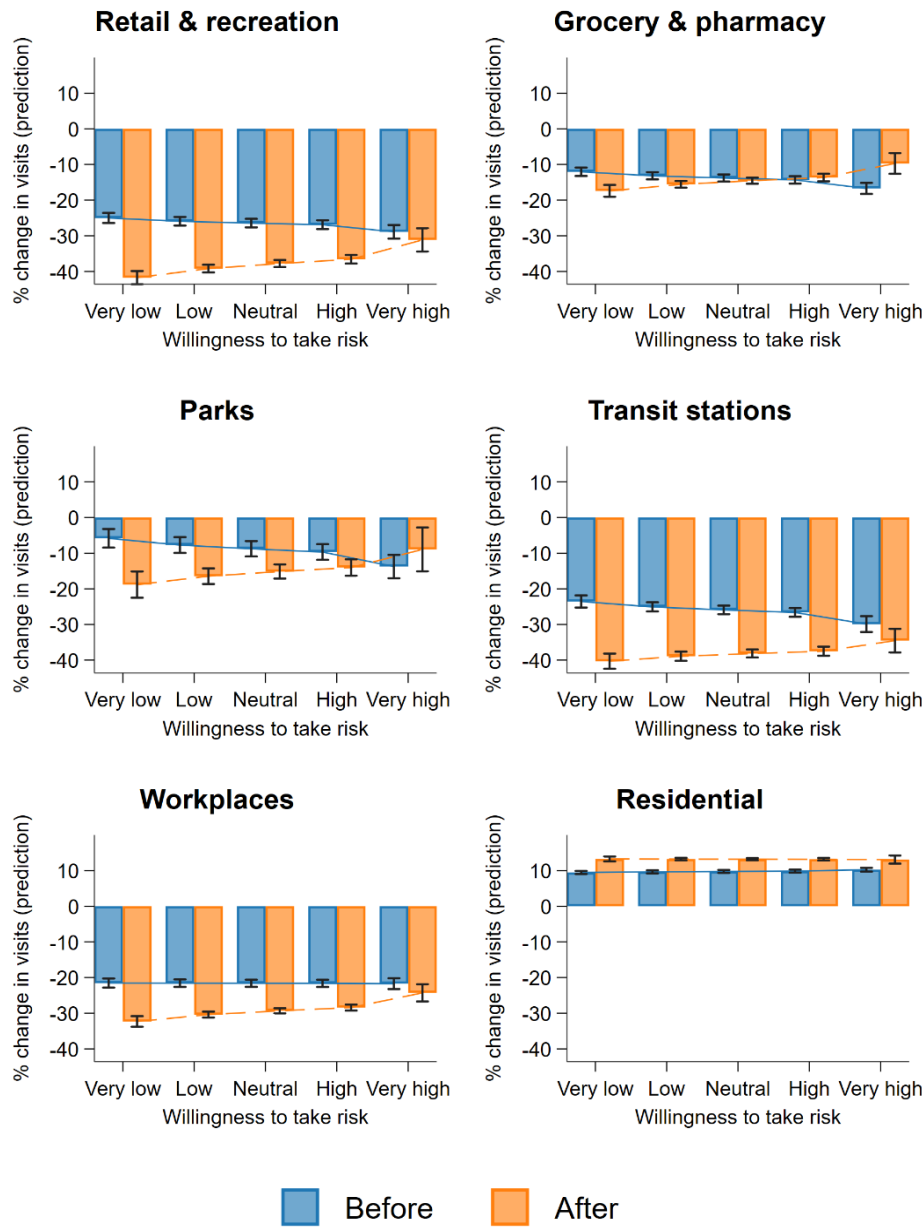

**Fig. 2 | Change in visits to six location categories predicted by average individual risk preference before and after pandemic declaration.** The six panels show the predicted percentage change in visit to locations classified as retail and recreation, grocery and pharmacy, parks, transit stations, workplaces, and residential, compared to the respective baseline values, before and after WHO declared COVID-19 as a pandemic on 11 March 2020, over average individual risk preference. Estimates are obtained from Table S2, for illustration, predicted changes are calculated over five points of the risk-taking variable (at the 1$^{st}$, 25$^{th}$, 50$^{th}$, 75$^{th}$, and 99$^{th}$ percentiles of the distribution), which we categorized into five levels of willingness to take risk: *very low*, *low*, *neutral*, *high*, and *very high*, respectively.

*Mobility patterns weekdays vs. weekends.* Next, we examine whether the tendency to change the frequency of visits to different localities during weekdays and weekends is mediated by risk attitude. As Fig. 3 shows, our earlier results are confirmed. Compared to weekdays, individuals on average further reduce their visits to places (compared to the same day of the week in the baseline period) classified as retail and recreation by 4.3pp ($\beta$=-4.272, s.e.=0.130, CI$_{95\%}$=[-4.527;-4.017], $P$<0.001; see Fig. 3), grocery and pharmacy by 3.9pp ($\beta$=-3.915, s.e.=0.164, CI$_{95\%}$=[-4.236;-3.594], $P$<0.001), parks by 4.4pp ($\beta$=-4.392, s.e.=0.450, CI$_{95\%}$=[-5.273;-3.511], $P$<0.001), and transit stations by 0.7pp ($\beta$=-0.723, s.e.=0.185, CI$_{95\%}$=[-1.085;-0.361], $P$<0.001), compared to the baseline. In contrast, the reduction in going to workplaces is larger during weekdays ($\beta$=8.342, s.e.=0.213, CI$_{95\%}$=[7.925;8.759], $P$<0.001), while individuals are more likely to stay home (places classified as residential) in general, the (percentage point) increase of staying home is higher during weekdays compared to weekends ($\beta$=-3.346, s.e.=0.109, CI$_{95\%}$=[-3.560;-3.131], $P$<0.001). The coefficients of the interaction terms provide strong evidence that regions with lower risk-tolerance have a larger reduction in mobility during weekends than in weekdays, compared to those who are more risk-tolerant (retail and recreation: $\beta$=2.011, s.e.=0.318, CI$_{95\%}$=[1.388;2.634], $P$<0.001; grocery and pharmacy: $\beta$=0.916, s.e.=0.411, CI$_{95\%}$=[0.110;1.722], $P$=0.026; parks: $\beta$=2.261, s.e.=1.015, CI$_{95\%}$=[0.272;4.250], $P$=0.026; transit stations: $\beta$=1.181, s.e.=0.502, CI$_{95\%}$=[0.197;2.165], $P$=0.019; workplaces: $\beta$=1.375, s.e.=0.506, CI$_{95\%}$=[0.384;2.367], $P$=0.007; residential: $\beta$=-0.789, s.e.=0.260, CI$_{95\%}$=[-1.298;-0.279], $P$=0.002). Results from robustness checks also confirm our findings (see Table S9 in SI Appendix).

    Moreover, we find that the mediation effect is more apparent after the declaration of pandemic, suggesting the effect manifests alongside severity. Specifically, we reran the analysis including the interaction between the risk preference-weekend mediation effect and pandemic declaration dummy (triple interaction term). We visualized the results in Fig. 4, showing the difference in average marginal effects of weekends (in contrast to weekdays) before and after the pandemic announcement, over levels of risk-taking (pre- and post-

declaration average marginal effects of weekends is shown in Fig. S1 and predicted change in mobility in Fig. S2). We find that the tendency to reduce going out during the weekends compared to weekdays increases significantly with the levels of risk-tolerance for all non-residential and work locations, particularly in the post-declaration period (retail recreation: $\beta=5.036$, s.e.=0.707, $CI_{95\%}=[3.651;6.421]$, $P<0.001$; grocery pharmacy: $\beta=4.273$, s.e.=0.698, $CI_{95\%}=[2.904;5.642]$, $P<0.001$; parks: $\beta=5.989$, s.e.=1.532, $CI_{95\%}=[2.985;8.993]$, $P<0.001$; transit stations: $\beta=4.697$, s.e.=0.884, $CI_{95\%}=[2.965;6.429]$, $P<0.001$). It can also be seen that regions with higher risk-taking attitude have a larger pre-post-declaration relative weekends-weekdays difference in mobility for workplaces ($\beta=4.008$, s.e.=0.665, $CI_{95\%}=[2.705;5.312]$, $P<0.001$) and residential places ($\beta=-1.397$, s.e.=0.290, $CI_{95\%}=[-1.966;-0.829]$, $P<0.001$). These results are highly robust to our checks (see Fig. S3 and Table S10 in SI Appendix).

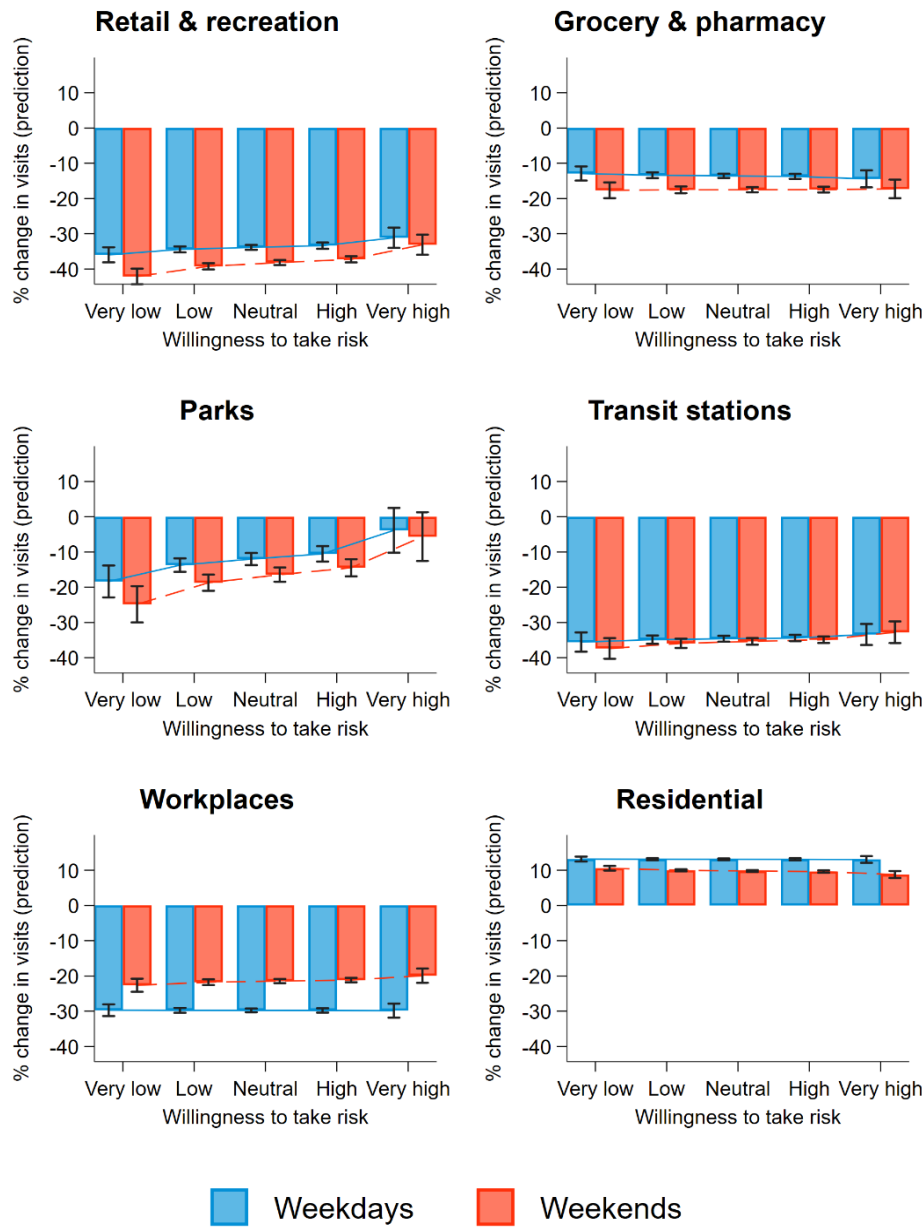

**Fig. 3 | Visitation pattern by weekdays and weekends over average individual risk preference.** The six panels show the predicted percentage change in visits to locations classified as retail and recreation, grocery and pharmacy, parks, transit stations, workplaces, and residential, compared to the respective baseline values in weekdays and weekends, over average individual risk-preference. Estimates are obtained from Table S3; for illustration, predicted changes are calculated over five points of the risk-taking variable (at the 1st, 25th, 50th, 75th, and 99th percentiles of the distribution), which we categorized into five levels of willingness to take risks: *very low*, *low*, *neutral*, *high*, and v*ery high*, respectively.

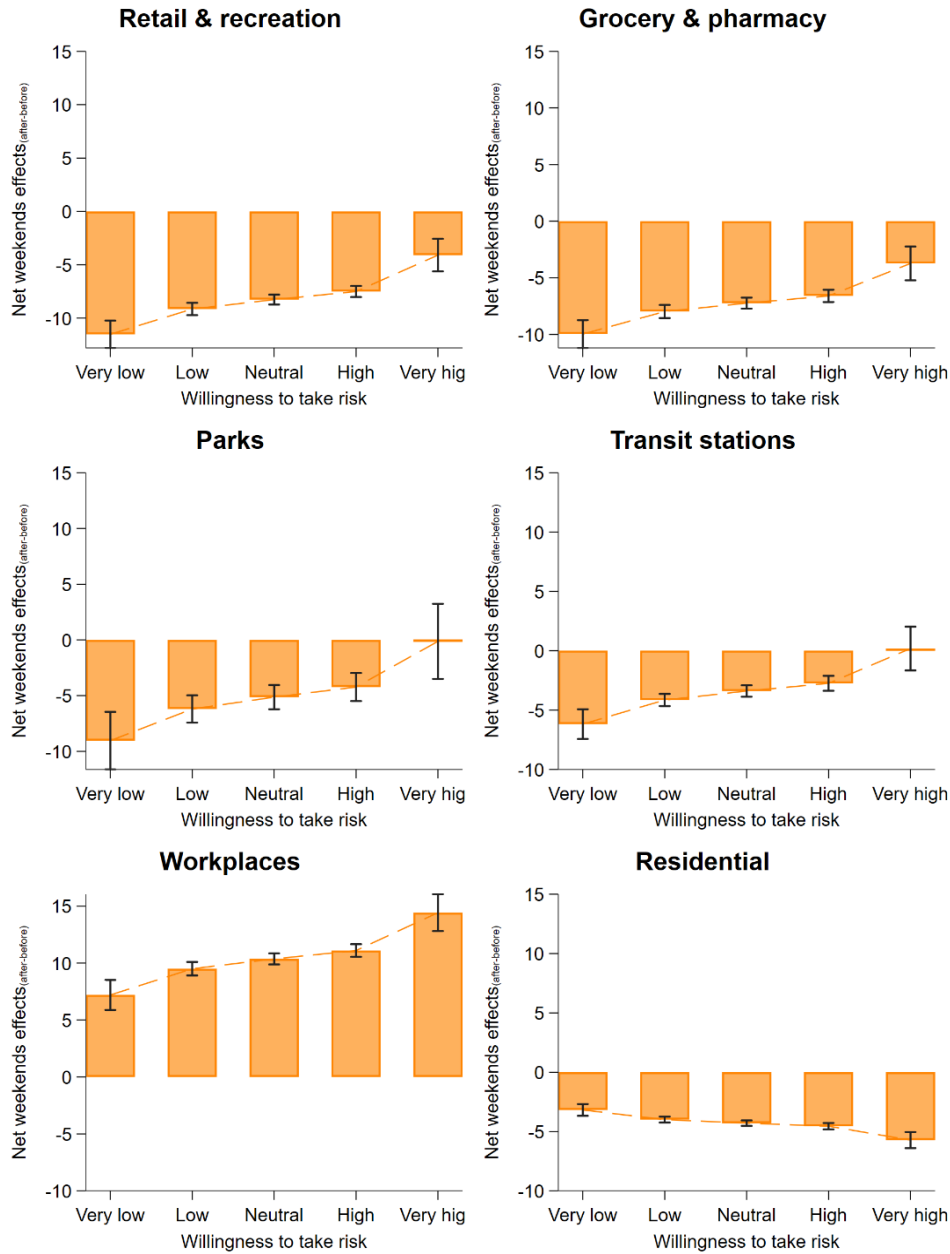

**Fig. 4 | Mediation from risk preference to change in weekends and weekdays visiting pattern is stronger after pandemic declaration.** The six panels show the difference in average marginal effects of weekends on visits to locations classified as retail and recreation, grocery and pharmacy, parks, transit stations, workplaces, and residential pre- and post-pandemic declaration periods, over risk-tolerance levels. Estimates are obtained from Table S4; for illustration, predicted changes are calculated over five points of the risk-taking variable (at the $1^{st}$, $25^{th}$, $50^{th}$, $75^{th}$, and $99^{th}$ percentiles of the distribution), which we categorized into five levels of willingness to take risks: *very low*, *low*, *neutral*, *high*, and *very high*, respectively.

*Actual risk.* Next, we examine the relationship between mobility changes, risk attitude, and proportion of elderly in the population to test if the relationship between mobility and risk is moderated by the share of population at higher risk of dying from COVID-19. We thus regressed change in mobility on willingness to take risk and share of population over 65 and the interaction between the two (see Fig. 5). We found that areas with a larger population at fatal risk (elderly) have larger cutback in going to grocery and pharmacy ($\beta$=-0.597, s.e.=0.164, CI$_{95\%}$=[-0.918;-0.277], $P$<0.001), transit stations ($\beta$=-0.364, s.e.=0.153, CI$_{95\%}$=[-0.664;-0.063], $P$=0.018), and workplaces ($\beta$=-0.447, s.e.=0.096, CI$_{95\%}$=[-0.636;-0.258], $P$<0.001), as well as a decrease in staying at home ($\beta$=-0.128, s.e.=0.049, CI$_{95\%}$=[-0.224;-0.033], $P$=0.009)[12], even though the size of the coefficients suggest the magnitude of the effect is quite small (e.g., with 1pp increase in share of over 65s in population, mobility change for staying home decreases by 0.1pp). While we did not find a significant (negative) change in mobility to retail and recreation ($\beta$=-0.226, s.e.=0.148, CI$_{95\%}$=[-0.517;0.065], $P$=0.128) and parks ($\beta$=-0.561, s.e.=0.484, CI$_{95\%}$=[-1.511;0.388], $P$=0.247) due to population risk level, results from the robustness checks show the negative effect is significant (see Table S11). Moreover, we found a significant interaction effects on mobility of retail and recreation ($\beta$=-0.388, s.e.=0.169, CI$_{95\%}$=[-0.719;-0.056], $P$=0.022) and residential ($\beta$ =-0.183, s.e.=0.050, CI$_{95\%}$=[-0.281;-0.086], $P$<0.001). This suggests that in areas with more risk-loving individuals and a larger proportion of population at risk, people seem to have further reduced going to retail and recreation places, however, regions with less risk-takers and larger older population increase their time staying home. Nonetheless, for mobility changes in other localities, the effect of risk-taking attitude does not seem to be moderated by the actual population risk factor.

---

[12] In our robustness checks excluding regions with censored mobility values due to insufficient data traffic, we found a weakly significant positive (at 10% level) effect.

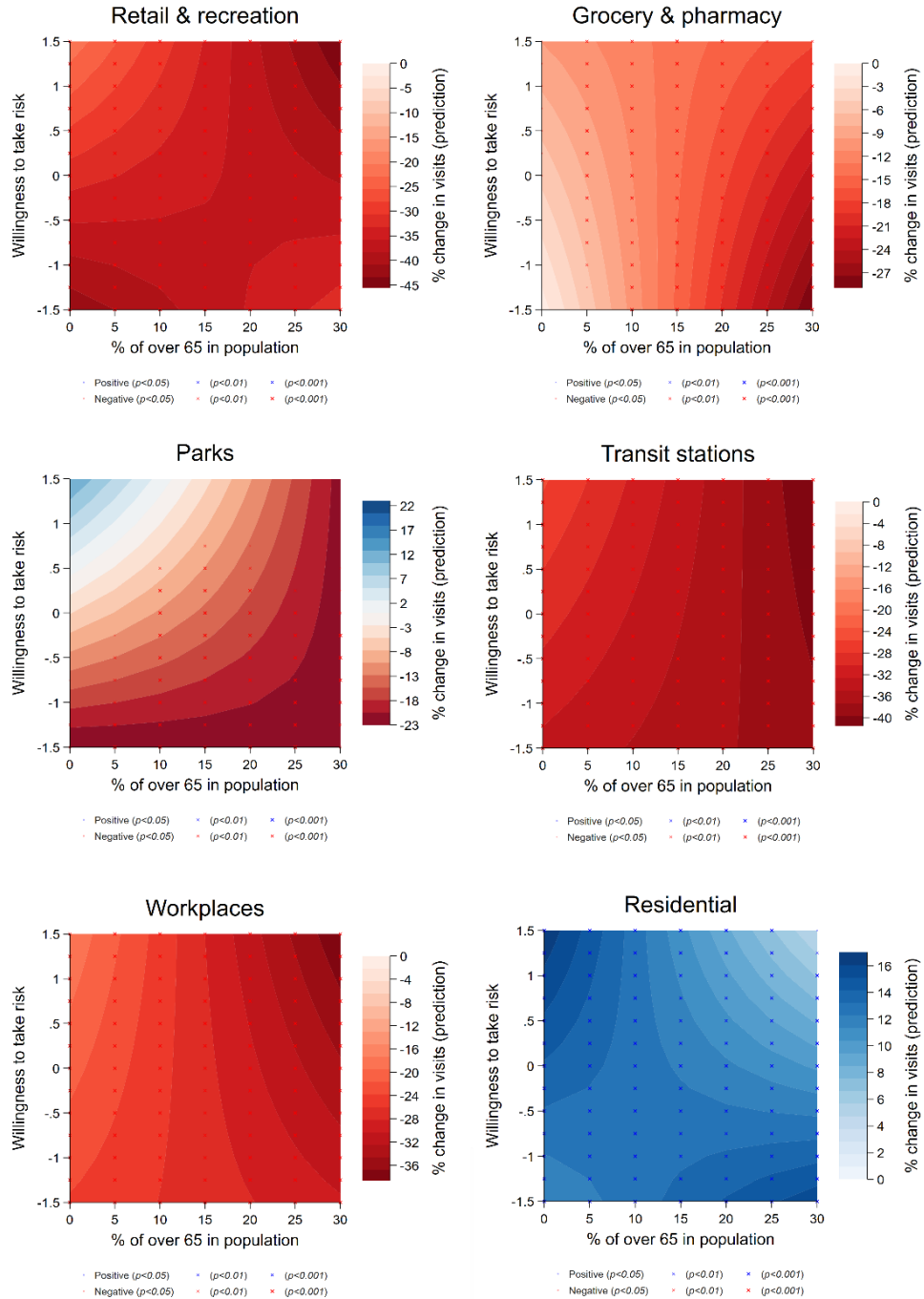

**Fig. 5 | Change of mobility patterns based on risk preference and share of population.** The six panels show the predicted change in visits to locations classified as retail and recreation, grocery and pharmacy, parks, transit stations, workplaces, and residential, over risk-tolerance levels and the proportion of over 65s in the population. Estimates are obtained from Table S5.

**Discussion**

As with Plato's cave, there are stark differences between how we perceive risk and the reality or the calculated level of risk, which can result in totally different behavioral outcomes. Risk attitudes clearly shape behavioral responses to pandemics. The actual health risks of the COVID-19 pandemic are (most likely) low for most groups apart from the elderly (40,41). In terms of mortality, the overall health consequences of Covid-19 could be similar to a pandemic influenza[13] (42). Nevertheless, risk attitude – rather than actual risk – influence real behavioral activity. Our results demonstrate the sharp shifts in the relation between behavioral activity and risk attitudes before and after declaration of COVID-19 as a pandemic, as well as shifts before and after the first related death was recorded. The first thing that becomes apparent is that behavior and our willingness to take on risks have both shifted dramatically since the baseline period in mid-February. At this stage, only three deaths were recorded outside mainland China[14] (one in Hong Kong, Japan, and the Philippines) and life was proceeding as normal. There was no imminent perception of a threat of the worldwide pandemic to come, reflected in the baseline reporting of behavior and the willingness to take risks. However, when we compare this to the first and second sample period, we observe mostly negative shifts in behavior (excluding residential) but a mixed set of reactions to risk. Several categories saw a substantial negative shift in visits, including Retail & Recreation, Transit Stations, and Workplaces; compared to the baseline, visiting behaviors had already started to drop off before the pandemic announcement but dropped off again afterward. During this first period, we can see that social distancing and work from home was starting to make an impact, as people stopped travelling to and from work

---

[13] Nevertheless, with immunization uncertainty and lack of vaccination and treatment, hospital can be overcrowded rapidly resulting overinflated infection and mortality rate.
[14] Figures taken from the Communicable Diseases Intelligence Report, Department of Health (Australia) https://www1.health.gov.au/internet/main/publishing.nsf/Content/1D03BCB527F40C8BCA258503000302EB/$File/covid_19_australia_epi_report_3_reporting_week_ending_1900_aedt_15_feb_2020.pdf

(especially through crowded transit stations) and also stopped engaging in non-essential retail shopping (therapy). After the pandemic was officially announced, we see a second wave of behavioral shifts as more people reduce their travel, shopping and more either lose their jobs or are in shutdown mode. However, we do observe an interesting shift in risk attitudes across these three categories as they all exhibit a slightly positive trend in the period before the announcement, but they all shift to a much stronger negative risk trend after the announcement. Given that 'flattening the curve' was the strategic focus for most governments, the social distancing message appears to have been received even prior to most lockdown measures. Conversely, Grocery & Pharmacy, Parks, and Residential had much smaller shifts both before and after the announcements when compared to the baseline. However, the shifts in Grocery & Pharmacy and Parks – while much smaller than the other categories – appear to undergo a large risk preference-mobility shift; that is, the first set of behavioral changes results in a positive sloping risk function that flipped into a negative sloped function after the pandemic announcement.

While seemingly at odds with expectations, one may want to consider what the announcement of the pandemic would have meant to most individuals. With a looming threat of lockdown and isolation, at this point individuals would have ramped up shopping to stock up for likely upcoming government lockdown. In addition, those with an affinity for the outdoors may have wanted to enjoy their parks and outdoor lifestyle as much as possible before it was banned. This is in line with the reported shifts in the number of visits, which while still negative overall, indicate that the change to number of visits is less negative than prior to the announcement. The odd one out is the Residential visits category; while small, we can still observe double increases in visitation numbers both pre and post the official pandemic announcement, and there is no change in the function representing the willingness to take risks.

When interpreting these statistics, we need to bear in mind the 'normal' weekly habits of people; that is, work during the week and undertake other activities/pastimes on the weekends. In order to ensure we capture the shift in behavior, we compare the weekday behaviors and risk attitudes to that of the weekends. Our result demonstrates that there are a few differences between weekdays and weekends, as one would expect that on weekends there are slightly more activities taking place other than work. Furthermore, we see little variance in the slopes of weekdays and weekends risk attitudes. The large negative shifts across all categories except workplaces after the official declaration, but much smaller variations between weekdays and weekends before the declaration, further supports the discussion above: that the behavior had already started to change well before the declaration of a pandemic, with many individuals starting to increase their weekend activities. However, after the pandemic was announced, a raft of measures that tried to limit the spread of the virus resulted in a very large change in most economies due to closure of businesses and job losses. This fundamental change in economic activities and loss of work left very little to differentiate weekends from weekdays for a large number of people, which is reflected in the large negative changes in the comparisons. Prior to the announcement, we see that the function on willingness to take risk is fairly flat or slightly downward sloping, but risk perceptions change significantly for all categories after the announcement. The most interesting changes are in Workplace and Residential, exhibiting a relatively large increase in the willingness to take high risks: this could be explained through people wanting to visit family and friends or the increased willingness to work despite the risk of infection.

In general, throughout our analysis we observe that less risk-tolerating regions more actively adjust their behavioral patterns in response to the pandemic. Risk seeking regions are less responsive to protective measures. Thus, the tendency towards being more careless or more cautious carries substantial behavioral implications that is also affected by different levels of opportunity costs, as evidenced by the weekend effect. Regional differences seem to matter, offering support for a "regional personality factor" in risk taking. As with individuals who allocate themselves to more risky professions there are regions that are likely more likely acting as "stunt persons", "fire-fighter", or "race-car driver regions".

Risk takers therefore seem to demonstrate a lower preference for their own and communal safety, as demonstrated that risk averse regions with higher percentages of 65+ people are more actively to increase social isolation by staying at home. Such behavioral differences due to risk preferences may indicate different levels of homeostatic responses. Risk aversion seems to promote a stronger fluctuation around a target level. For example, if you are driving on a motorway and it starts to rain or snow, what do you do? Our result would imply that risk averse individuals may be more likely to slow down to reduce the likelihood of having an accident. Risk averse individuals have a higher need for risk compensation. Thus, the level of risk at which a person feels best is maintained homeostatically in relation to factors such as emotional or physiological experiences (19).

Overall, the lack of adjustment among risk taking regions is interesting, as many settings that explore risk taking behavior are connected to the possibility of attracting social fame and praise, financial gains, or other potential positive outcomes. In our setting, the risks are strongly attached to the loss of their own and other's health or life without achieving major gains, although positive utility gains also arise from not restricting one's usual activities. It seems like the risk takers are more "pathologically" stable during such environmentally challenging circumstances. It is almost as if risk taking regions are more determined to maintain settings as activity-oriented, while risk averse regions are more goal-oriented in achieving social distancing.

The current analysis is interesting, as a large number of studies exploring the implications of risk are based on cross-sectional samples or between-subject designs in laboratory settings. In this case, the danger is more prolonged, lasting over several weeks or months, compared with other risk situations such as driving a car. Automatic or response "scripts" become less relevant as individuals have the chance to think about their actions and adjust their behavior accordingly. Strategic, tactical, or operational factors become more dominant while perceptual, emotional, and motivational factors remain active. In addition, individuals do not face a single "either-or" decision but are required to constantly evaluate their choices to go out or stay at home. Thus, cognitive reevaluation is a core feature in our setting, and is based on dynamic feedback loops. Risk loving regions are also less likely to adjust their behavior based on external stimulus such as the WHO announcement of classifying COVID-19 as a pandemic.

A core limitation is that we are only able to explore human behavior at the regional and not individual level. Studies that use individual data could focus in more detail on individual differences such as age or gender or differences in affective reactions or perceived locus of control and could try to disentangle perceptions (risk preferences) partly from actual risk as statistics provide detailed information on the actual age risk profile. Such a study would provide a better understanding of habit changes, as well as potentially reveal motivational reasons for behavioral changes or behavioral stickiness. To reduce levels of uncertainty or ambiguity, individuals will try to gain control over a situation or they will change their preferences to better the fit the situation, and thus try to gain control in a secondary way (19). Other psychological factors such as overconfidence may also matter. In addition, we do not have information about the actual level of social mobility in the baseline time period. If that information were available, one could argue that those who had the highest levels of mobility prior to the lockdown have had the largest relative loss; we should therefore observe this group exhibiting the most risk seeking behavior and breaking the lockdown rules. On the other hand, those who previously had the least amount of social mobility have in relative terms only suffered a small loss – and should be much less likely to break the lockdown rules. However, this may adjust over time, as individuals habituate to the changes and reset their reference points. This fits nicely into the suggestion that "a person who has not made peace with his losses is likely to accept gambles that would be unacceptable to him otherwise" (29, p. 287), which is consistent with risk preference changes in a disaster situation (10).

Risk is a fascinating topic as we have two forces in place. Based on evolutionary theory, people are risk-inclined but also control-inclined. Risk taking is necessary to cope with environmental changes and the constant level of uncertainty and danger. On other hand, control of the environment is required to reduce risks that go beyond the desired levels or that may pose danger to one's survival (19).

The pandemic declaration caused a fundamental shift in behavior, independently of government lockdown measures. Future studies could explore in more detail how information dissemination and media reporting are connected to behavioral responses and the level of risk taking within regions. Removal of the lockdown policies is likely to be undertaken cautiously and slowly rather than via one large change. It is unclear at this stage how changes – particularly among the risk averse regions – have already led to new habit formation that will not readjust to previously normal settings. Future studies will provide more insights into such a question.

**Material and Methods**

**Mobility**. We obtained the mobility measures on country and regional level from the COVID-19 Community Mobility Reports (43), assessed on 16 May 2020[15]. The dataset consists of six location-specific mobility measures for 132 countries between 15 February 2020 and 09 May 2020. For 51 out of the 131 countries, the mobility measures are also available at the regional level. For the United States, both state and county level is available, although we did not include county level in our analysis as risk preference is not available at the county level. The resulting number of sub-national regions included is 1,207. Based on anonymized and aggregated data from Google users who have opted in to their Location History service, each mobility measure records the percent change in visits and length of stay to places classified as *Retail & Recreation*, *Grocery & Pharmacy*, *Parks*, *Transit Stations*, *Workplaces*, and *Residential* within the geographic area. The percent change is compared to the median value of the same day of the week between 3 January and 6 February 2020. For privacy reasons, Google censored values if the traffic volume is not high enough to ensure anonymity. While the median number of censored values for each mobility measure is zero, about 48% (*n*=619) of regions have at least one censored value for any of the six mobility variables on any given day in the sample period. To ensure our results were not caused by the unbalanced sample due to censored values, we reran our results by excluding regions with various thresholds of daily values censored, finding that the results remain highly robust to all exclusions (see Table S7 to S11 in SI Appendix).

**Risk attitude**. We obtain the measure of risk preference from the globally representative Global Preferences Survey collected in 2012 using the Gallup World Poll (36, 37), which is aggregated into the country (*n*=76) and regional (*n*=1,126) level. Risk preferences of the respondents were elicited through a qualitative question (self-rated perceived risk preference on a 11-point scale) and a set of quantitative questions using the staircase method, where respondents were asked to choose between varying sure payments and a fixed lottery, in which the individual could win x with some probability *p* or zero. The responses from the two questions were combined (with roughly equal weights) to produce the overall individual risk preference measure (37). For subnational regions where both mobility measures and risk preference measures are available at the region levels, we employed the regional aggregated values (average of standardized values at the individual level), otherwise country aggregated values were used.

---

[15] Before Google officially release of the mobility data file on 15 April 2020, an earlier version of the data was obtained from (44, https://osf.io/rzd8k/), based on values extracted from each PDF file of the Mobility Reports using WebPlotDigitizer (45).

**Covid-19 cases and deaths statistics and government response indicators**. Country-level statistics on the daily number of cases and deaths were sourced from the European Centre for Disease Prevention and Control (ECDC). Together with the set of indicators on government responses, these data were obtained from the Oxford COVID-19 Government Response Tracker (OxCGRT) (39, assessed on 18 May 2020), available for 167 countries recorded daily from 01 January 2020. Out of the 17 response indicators available from the OxCGRT, we take seven indicators on policies regarding social isolation and confinement, including school, workplace, and public transport closures, public events cancellation, stay at home requirements, and gatherings and internal movement restrictions. Each indicator has various levels of response, from no measures taken to recommendation and implementation of the policy, recorded on ordinal scale. For example, workplace closure is classified into four levels (1 – no measures; 2 – recommend work from home; 3 – require closing for some categories or sectors of workers; and 4 – require closing for all-but-essential workplaces)[16]. We dichotomously coded each response to be included in our regression analysis. OxCGRT also records if the policy is applied nationwide; for robustness checks, we recode the each response indicators as no measures taken if policy is targeted to a specific geographical region (see SI Appendix).

---

[16] See https://github.com/OxCGRT/covid-policy-tracker/blob/master/documentation/codebook.md and (39) for more detailed definition of each containment and closure policies variable.

**Control variables**. *Population*. Population density (people per squared km of land area), percentage of urban population, share of population ages 65 were obtained from the World Development Indicators (46) and are available at the country level. We also collected the average household size (average number of usual residents per household) from the Database on Household Size and Composition 2019 (47). *Economic indicators.* We also obtained the latest available estimates (as of May 2020) of unemployment rate (% of total labor force), based on estimates from the International Labour Organization, and per capita GDP (2010 US$ constant, in natural log form) from (46). *Weekend*. We employ the definition of working week across countries according to (48). *Daily average temperature*. Temperature data is obtained from the GHCN (Global Historical Climatology Network)-Daily database (49, 50), assessed on 19 May 2020. For each region, the daily average temperature (in tenths of degree Celsius (°C)) was calculated from taking the mean[17] of the average temperature recorded from all weather stations located within 50km from the centroid of the region.

**Combining datasets**. To join datasets together for our analysis, we use regions defined in the Google Mobility dataset as our point of reference. In general, for regions with mobility measures but not from another dataset (i.e., risk attitude or average daily temperature is unavailable for that region), we employ its country values. The resulting number of countries in our final sample is 58, after merging all variables used in this study, with a total of 776 subnational regions from 33 countries (see Table S6 in SI Appendix). The total number of region-day observations ranges from 58,284 to 67,073, depending on the availability of mobility measures.

---

[17] Using median does not change our results.

**Analyses**. To examine the main question of how mobility patterns during the COVID-19 outbreak change according to risk attitude, we analyzed the data using random-effects linear model. Standard errors are clustered on the smallest geographic unit in each regression. Data and codes used in this study can be found on Open Science Framework (https://osf.io/7bxqp/).

## Acknowledgments

We received no specific funding for this work. We thank D. Johnston for helpful feedback.

## Author Contributions

A.S, and B.T. conceived the idea. H.F.C., A.S, and B.T. designed the research. H.F.C. and A.S. acquired the data. H.F.C. and B.T. performed the analyses and draft the paper. H.F.C. performed the visualization of the results and managed the Supplementary Information. H.F.C., A.S, D.Sa., D.St., and B.T. wrote, read, and revised the manuscript. All authors provided critical intellectual contributions into aspects of this study.

## Competing Interests statement

The authors declare no competing interests.

**Supplementary Information for:**

**Risk Attitudes and Human Mobility during the COVID-19 Pandemic**

## Supplementary Results

**Robustness Checks.**

This section presents the checks for robustness of our results, which are shown in Table S7 to S11 for the six sets of regressions conducted in the main text, respectively. The first two checks concern including regions with censored mobility value in the sample of the analysis. We impose two restrictions on sample inclusion 1) regions with at least one censored value for the outcome mobility measures are excluded from the corresponding regression and 2) a more restrictive criteria with regions at least one censored value for any of the outcome mobility measures are excluded from the analysis. The first criteria excluded number of regions ranging from 54 (Workplace) to 217 (Residential), depending on the outcome mobility measure used while the second criteria reduce the number of regions to 484. The third check concerns if estimates are sensitive to whether government response is general by recoding indicators as no measures taken if the movement restrictions (or recommendation of restrictions) were not applied countrywide. In general, our main findings are robust to all three checks.

For the overall risk-mobility relationship (comparing estimates from Table S7 to Table S1), imposing sample exclusions increases the strength of the relationship for all mobility measures except for transit station under the first exclusion rule (first exclusion rule: retail & recreation: $\beta=2.219$, s.e.=1.183, $CI_{95\%}=[-0.099;4.537]$, $P=0.061$; grocery & pharmacy: $\beta=-0.325$, s.e.=1.006, $CI_{95\%}=[-2.298;1.647]$, $P=0.746$; parks: $\beta=8.019$, s.e.=2.535, $CI_{95\%}=[3.050;12.987]$, $P=0.002$; transit stations: $\beta=1.306$, s.e.=1.352, $CI_{95\%}=[-1.344;3.956]$, $P=0.334$; workplaces: $\beta=0.761$, s.e.=0.854, $CI_{95\%}=[-0.913;2.434]$, $P=0.373$; residential: $\beta=0.089$, s.e.=0.413, $CI_{95\%}=[-0.721;0.900]$, $P=0.829$; second exclusion rule: retail & recreation: $\beta=2.551$, s.e.=1.074, $CI_{95\%}=[0.447;4.656]$, $P=0.017$; grocery & pharmacy: $\beta=-0.279$, s.e.=0.905, $CI_{95\%}=[-2.052;1.494]$, $P=0.758$; parks: $\beta=9.463$, s.e.=2.589, $CI_{95\%}=[4.388;14.539]$, $P<0.001$; transit stations: $\beta=1.865$, s.e.=1.186, $CI_{95\%}=[-0.459;4.189]$, $P=0.116$; workplaces: $\beta=-0.246$, s.e.=0.773, $CI_{95\%}=[-1.761;1.269]$, $P=0.751$; residential: $\beta=0.251$, s.e.=0.427, $CI_{95\%}=[-0.587;1.088]$, $P=0.557$). Transforming the government response indicators slight reduce the size of the coefficients while leaving the statistical significance unchanged (retail & recreation: $\beta=3.724$, s.e.=1.225, $CI_{95\%}=[1.324;6.124]$, $P=0.002$; grocery & pharmacy: $\beta=-0.247$, s.e.=1.138, $CI_{95\%}=[-2.477;1.984]$, $P=0.828$; parks: $\beta=6.070$,

s.e.=2.455, CI$_{95\%}$=[1.258;10.882], $P$=0.013; transit stations: $\beta$=1.597, s.e.=1.597, CI$_{95\%}$=[-1.534;4.727], $P$=0.317; workplaces: $\beta$=0.581, s.e.=0.940, CI$_{95\%}$=[-1.261;2.424], $P$=0.536; residential: $\beta$=-0.481, s.e.=0.374, CI$_{95\%}$=[-1.213;0.252], $P$=0.199). The coefficient estimates for our main control variables (*pandemic declaration* and *weekends*) are also close to those found in the main results, apart from *% population 65+*, where the negative effects on mobility change to non-residential places are more prominent in the restricted samples (checks 1 and 2).

For the declaration moderator effect on the risk-mobility relationship, the results (coefficients of the declaration x risk preference term) remain highly robust except for residential, where statistically significance is drop when regions with censored values were removed (first exclusion rule: Retail & recreation: $\beta$=4.965, s.e.=1.210, CI$_{95\%}$=[2.593;7.337], $P$<0.001; Grocery & pharmacy: $\beta$=6.265, s.e.=1.075, CI$_{95\%}$=[4.157;8.372], $P$<0.001; *Parks*: $\beta$=10.471, s.e.=2.477, CI$_{95\%}$=[5.617;15.325], $P$<0.001; Transit stations: $\beta$=6.919, s.e.=1.514, CI$_{95\%}$=[3.952;9.887], $P$<0.001; Workplaces: $\beta$=3.768, s.e.=0.880, CI$_{95\%}$=[2.042;5.494], $P$<0.001; Residential: $\beta$=-0.216, s.e.=0.515, CI$_{95\%}$=[-1.225;0.793], $P$=0.675; second exclusion rule: Retail & recreation: $\beta$=4.702, s.e.=1.388, CI$_{95\%}$=[1.981;7.422], $P$<0.001; Grocery & pharmacy: $\beta$=4.146, s.e.=1.190, CI$_{95\%}$=[1.815;6.478], $P$<0.001; *Parks*: $\beta$=11.653, s.e.=2.695, CI$_{95\%}$=[6.372;16.935], $P$<0.001; Transit stations: $\beta$=4.885, s.e.=1.698, CI$_{95\%}$=[1.556;8.213], $P$=0.004; Workplaces: $\beta$=1.504, s.e.=0.995, CI$_{95\%}$=[-0.446;3.453], $P$=0.131; Residential: $\beta$=-0.092, s.e.=0.548, CI$_{95\%}$=[-1.166;0.983], $P$=0.867; government response indicators transformed: Retail & recreation: $\beta$=2.983, s.e.=1.083, CI$_{95\%}$=[0.860;5.107], $P$=0.006; Grocery & pharmacy: $\beta$=2.926, s.e.=1.025, CI$_{95\%}$=[0.917;4.936], $P$=0.004; *Parks*: $\beta$=7.071, s.e.=2.382, CI$_{95\%}$=[2.403;11.740], $P$=0.003; Transit stations: $\beta$=2.934, s.e.=1.322, CI$_{95\%}$=[0.343;5.526], $P$=0.026; Workplaces: $\beta$=0.284, s.e.=0.923, CI$_{95\%}$=[-1.524;2.092], $P$=0.758; Residential: $\beta$=1.065, s.e.=0.443, CI$_{95\%}$=[0.197;1.934], $P$=0.016).

For the weekend x risk-taking interaction term, the coefficients for retail and recreation remain statistically significant in all three checks (first exclusion rule: $\beta$=1.622, s.e.=0.308, CI$_{95\%}$=[1.018;2.225], $P$<0.001; second exclusion rule: $\beta$=2.038, s.e.=0.349, CI$_{95\%}$=[1.354;2.722], $P$<0.001; government response indicators transformed: $\beta$=1.464, s.e.=0.331, CI$_{95\%}$=[0.814;2.113], $P$<0.001) as well as for parks (first exclusion rule: $\beta$=3.441, s.e.=1.092, CI$_{95\%}$=[1.301;5.581], $P$=0.002; second exclusion rule: $\beta$=4.474, s.e.=1.306, CI$_{95\%}$=[1.914;7.035], $P$<0.001; government response indicators transformed: $\beta$=1.997,

s.e.=1.026, CI$_{95\%}$=[-0.015;4.009], *P*=0.052), transit stations (first exclusion rule: *β*=1.251, s.e.=0.541, CI$_{95\%}$=[0.192;2.311], *P*=0.021; second exclusion rule: *β*=1.964, s.e.=0.585, CI$_{95\%}$=[0.817;3.112], *P*<0.001; government response indicators transformed: *β*=0.914, s.e.=0.494, CI$_{95\%}$=[-0.053;1.882], *P*=0.064), workplaces (first exclusion rule: *β*=1.316, s.e.=0.511, CI$_{95\%}$=[0.314;2.317], *P*=0.010; second exclusion rule: *β*=2.065, s.e.=0.635, CI$_{95\%}$=[0.820;3.310], *P*=0.001; government response indicators transformed: *β*=1.115, s.e.=0.494, CI$_{95\%}$=[0.147;2.083], *P*=0.024), and residential area (first exclusion rule: *β*=-1.141, s.e.=0.305, CI$_{95\%}$=[-1.739;-0.543], *P*<0.001; second exclusion rule: *β*=-1.255, s.e.=0.321, CI$_{95\%}$=[-1.884;-0.626], *P*<0.001; government response indicators transformed: *β*=-0.724, s.e.=0.259, CI$_{95\%}$=[-1.232;-0.215], *P*=0.005). Transforming government response indicators rendered the significance of the interaction effects for going to grocery and pharmacy (*β*=0.466, s.e.=0.420, CI$_{95\%}$=[-0.356;1.289], *P*=0.267; first exclusion rule: *β*=0.697, s.e.=0.406, CI$_{95\%}$=[-0.099;1.494], *P*=0.086; second exclusion rule: *β*=0.800, s.e.=0.454, CI$_{95\%}$=[-0.091;1.691], *P*=0.078). This suggests that the tendency to further reduce mobility on the weekends than during the week for low risk-tolerance regions (as compared to high risk-tolerance regions) is evident before pandemic declaration. Moreover, we see that the results with triple interactions between risk preference, weekend, and pandemic declaration resembles to that in the main text, albeit for regions with very high risk preference, the pre- and post-declaration difference in the weekend reduction in mobility is less precisely estimated in the second sample restriction, in particular for retail and recreation, grocery and pharmacy, and parks.

Lastly, we found some of the estimates of the risk preference-risk pool interaction terms is similar to that in the main analysis. For retail & recreation, the first exclusion rule (*β*=-0.365, s.e.=0.185, CI$_{95\%}$=[-0.729;-0.002], *P*=0.049) and second exclusion rule (*β*=-0.550, s.e.=0.176, CI$_{95\%}$=[-0.894;-0.205], *P*=0.002) both result in significant interaction terms, while transforming the government response indicators, the significance disappeared (*β*=-0.046, s.e.=0.163, CI$_{95\%}$=[-0.365;0.273], *P*=0.777). For residential area, the negative interaction terms is highly robust (first exclusion rule: *β*=-0.167, s.e.=0.060, CI$_{95\%}$=[-0.284;-0.050], *P*=0.005; second exclusion rule: *β*=-0.189, s.e.=0.063, CI$_{95\%}$=[-0.313;-0.066], *P*=0.003; government response indicators transformed: *β*=-0.268, s.e.=0.045, CI$_{95\%}$=[-0.355;-0.181], *P*<0.001). For other localities, the coefficient of the interaction term is consistently not statistically significant.

# Supplementary Figures

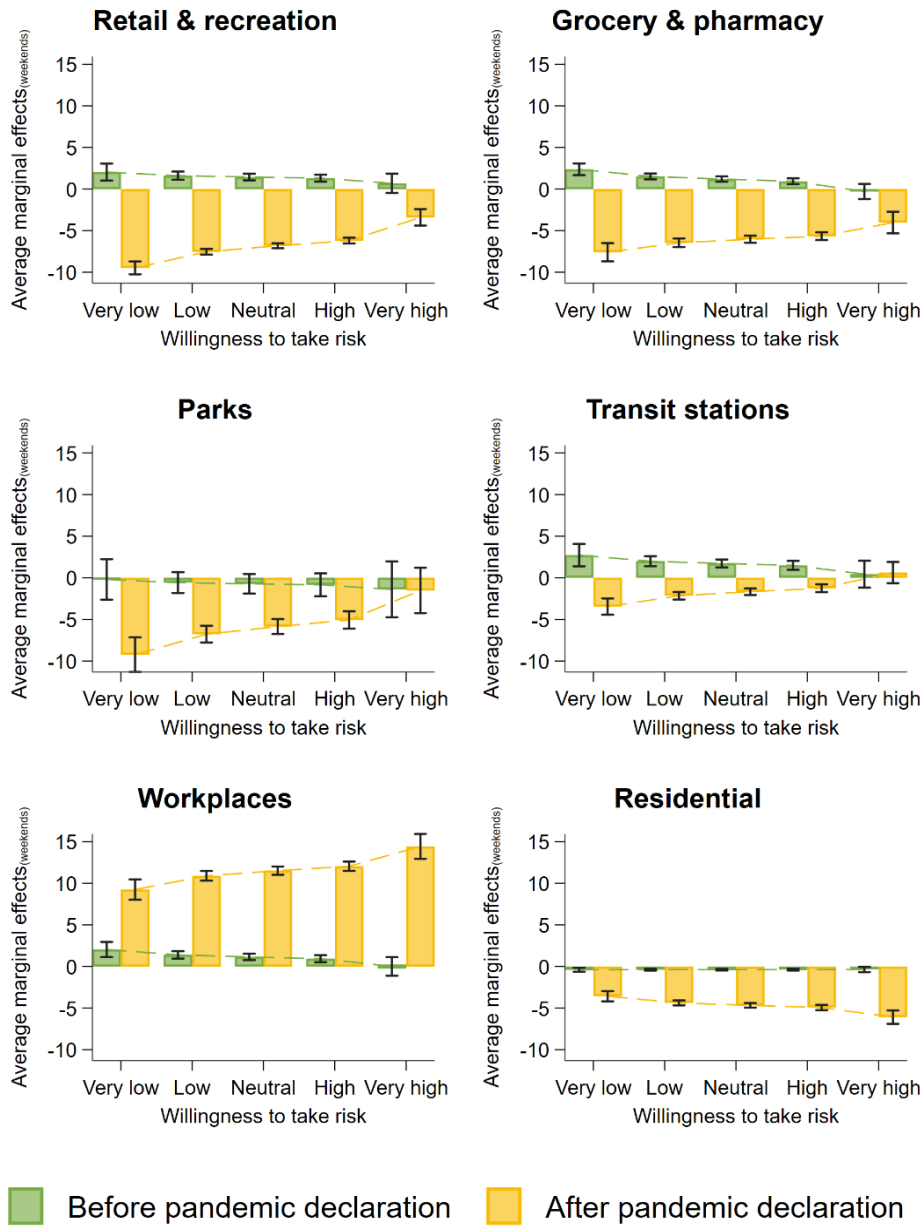

**Fig. S1 | Average marginal effects of weekends on mobility changes over risk attitudes, before and after pandemic declaration.**

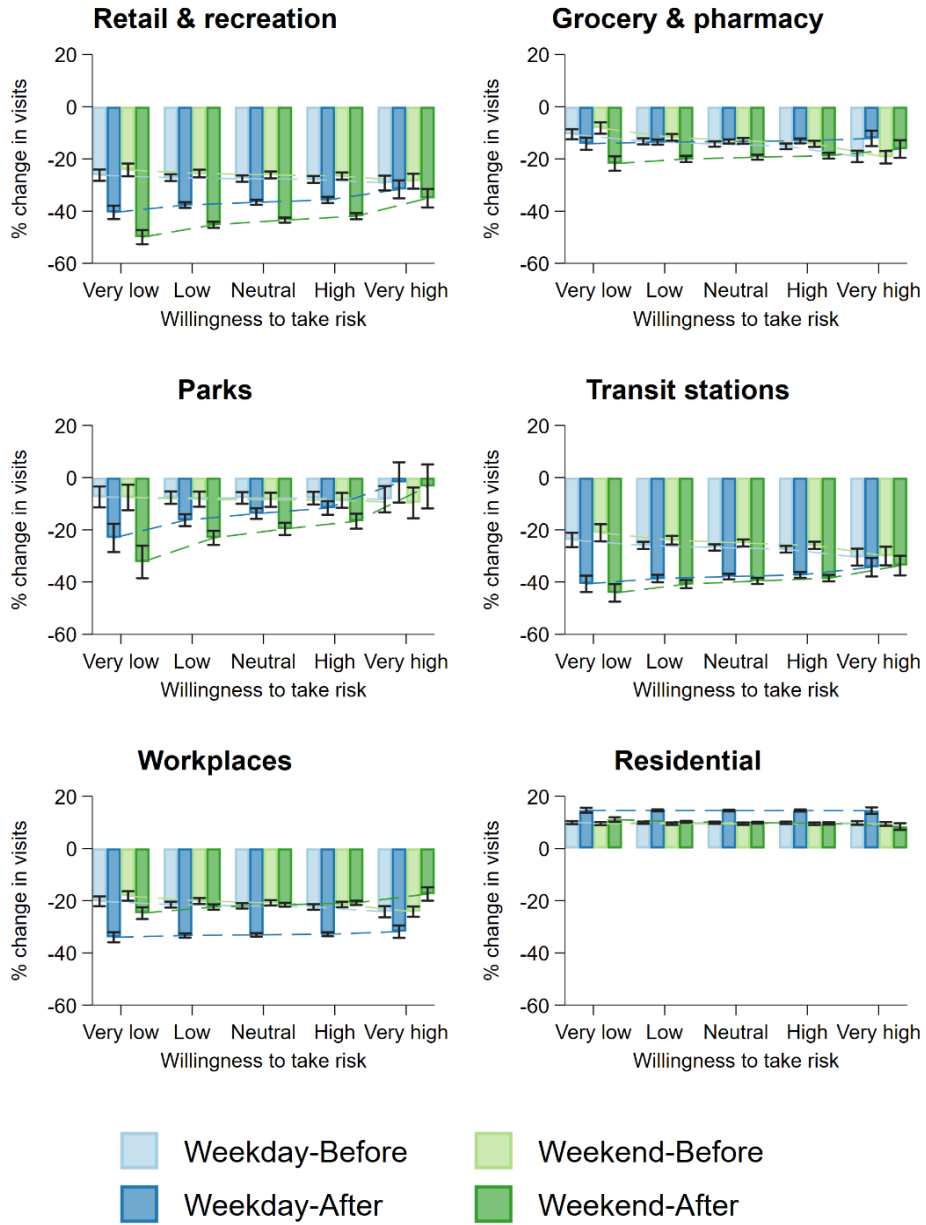

**Fig. S2 | Predicted change in mobility on weekdays and weekends and before and after pandemic declaration, over risk attitudes.**

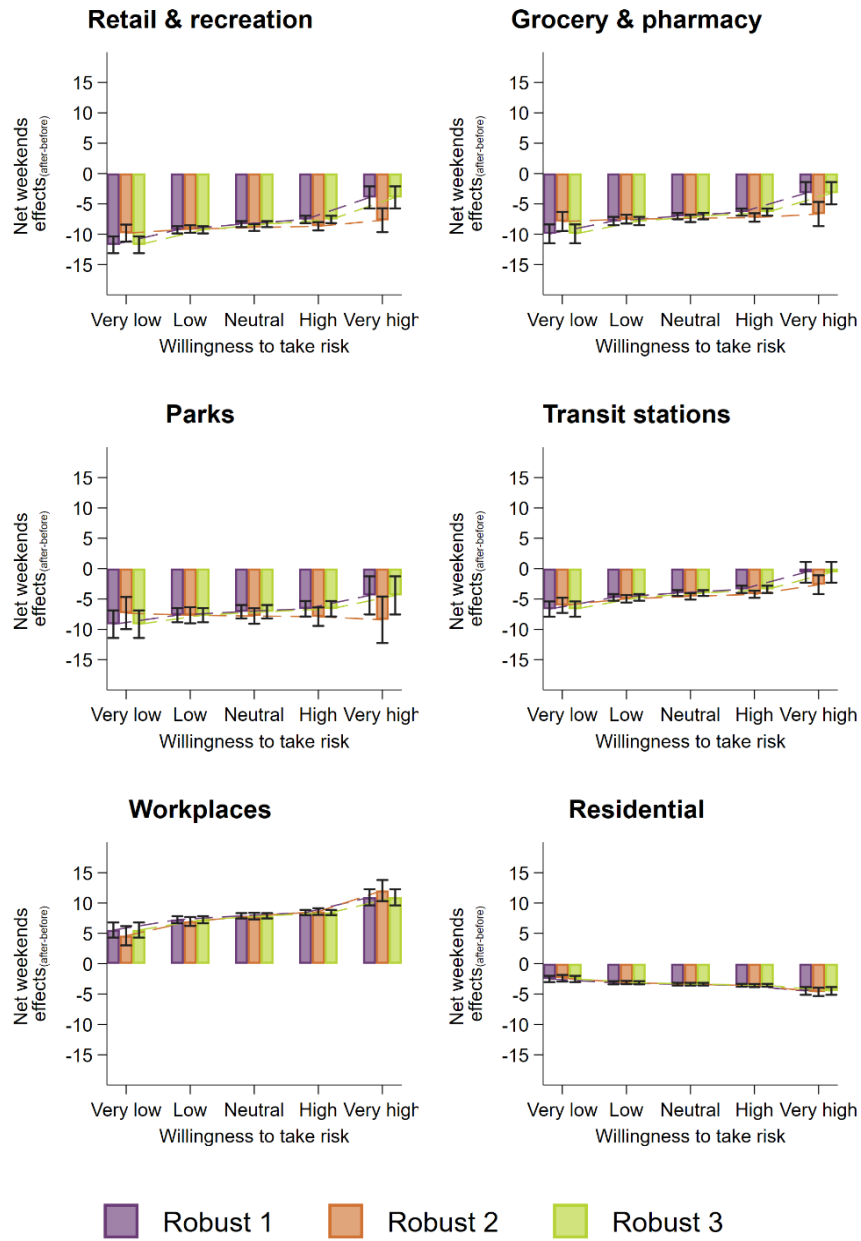

**Fig. S3 | Robustness checks on mediation from risk preference to change in weekends and weekdays visiting pattern before and after pandemic declaration.** Robust 1 = regions with at least one censored values on the outcome mobility measures excluded. Robust 2 = regions with at least one censored values on any mobility measures excluded. Robust 3 = government response indicators recoded as no measures taken if policy is not applied countrywide.

# Supplementary Tables

## Table S1 | Risk attitude and human mobility and during COVID-19

|  | Retail & recreation | Grocery & pharmacy | Parks | Transit stations | Workplaces | Residential |
|---|---|---|---|---|---|---|
| Risk-taking | 2.87* | -0.48 | 7.67** | 1.35 | 0.31 | -0.24 |
|  | (1.180) | (1.060) | (2.577) | (1.350) | (0.848) | (0.374) |
| Pandemic declaration | -11.8*** | -1.42* | -8.38*** | -12.6*** | -8.38*** | 3.62*** |
|  | (0.879) | (0.711) | (1.373) | (0.828) | (0.649) | (0.287) |
| Weekends | -4.39*** | -3.97*** | -4.54*** | -0.79*** | 8.28*** | -3.28*** |
|  | (0.132) | (0.167) | (0.446) | (0.189) | (0.215) | (0.110) |
| Days after first death | 0.042† | 0.13*** | 0.056 | -0.12*** | 0.046* | 0.017* |
|  | (0.0216) | (0.0171) | (0.0352) | (0.0242) | (0.0190) | (0.00831) |
| ln(# confirmed cases+1) | -2.32*** | -1.86*** | 1.69† | 1.72*** | -1.53*** | 0.13 |
|  | (0.466) | (0.395) | (0.881) | (0.481) | (0.358) | (0.165) |
| School |  |  |  |  |  |  |
| *Recommend closing* | -7.61*** | 0.98 | -30.9*** | 2.17 | 3.95*** | 2.50*** |
|  | (2.103) | (1.087) | (4.999) | (1.614) | (1.037) | (0.582) |
| *Require closing (some)* | 6.68*** | 1.08 | 15.6*** | 5.18† | 0.11 | -1.43*** |
|  | (1.832) | (1.670) | (3.185) | (2.728) | (1.242) | (0.284) |
| *Require closing* | -6.65*** | -5.53*** | -4.26*** | -2.75*** | -5.92*** | 2.49*** |
|  | (0.764) | (0.600) | (1.151) | (0.736) | (0.598) | (0.233) |
| Workplace closing |  |  |  |  |  |  |
| *Recommend closing* | -1.69† | 5.47*** | 8.02*** | -3.09*** | 0.90 | -0.44 |
|  | (0.907) | (0.904) | (1.346) | (0.930) | (0.845) | (0.302) |
| *Require closing (some)* | -23.5*** | -7.33*** | -3.29† | -16.0*** | -11.9*** | 4.03*** |
|  | (1.224) | (1.097) | (1.825) | (1.291) | (0.881) | (0.371) |
| *Require closing* | -16.8*** | -4.50*** | -5.76* | -17.5*** | -9.79*** | 4.38*** |
|  | (1.334) | (1.183) | (2.416) | (1.367) | (0.956) | (0.432) |
| Public events |  |  |  |  |  |  |
| *Recommend cancelling* | -0.69 | 2.56* | -5.61*** | -4.37** | -2.34** | 1.87*** |
|  | (1.180) | (1.017) | (1.322) | (1.504) | (0.793) | (0.233) |
| *Require cancelling* | -5.04*** | -0.42 | -3.91* | -3.78*** | -4.06*** | 2.45*** |
|  | (0.716) | (0.687) | (1.652) | (0.869) | (0.674) | (0.229) |
| Restrictions on gatherings |  |  |  |  |  |  |
| *Above 1000 people* | 10.5*** | 8.94*** | 1.69 | 4.17*** | 6.68*** | -3.87*** |
|  | (1.003) | (1.091) | (2.010) | (1.258) | (0.883) | (0.320) |
| *101-1000 people* | 4.97*** | 8.12*** | 2.58 | 2.30† | 5.58*** | -2.41*** |
|  | (1.245) | (1.077) | (2.041) | (1.366) | (0.861) | (0.323) |
| *11-100 people* | -1.29 | -1.06 | -5.11* | -3.58* | -4.64*** | 0.011 |
|  | (1.398) | (1.245) | (2.456) | (1.504) | (1.083) | (0.430) |
| *10 people or less* | -1.68 | -0.59 | -3.62† | -3.53** | -2.67** | -0.21 |
|  | (1.192) | (1.065) | (1.877) | (1.257) | (0.915) | (0.357) |
| Public transport |  |  |  |  |  |  |
| *Recommend closing* | -8.52*** | -9.87*** | -10.4*** | -8.77*** | -10.8*** | 4.31*** |
|  | (0.795) | (0.702) | (1.559) | (0.770) | (0.613) | (0.253) |
| *Require closing* | -7.44*** | -11.6*** | -18.3*** | -6.73*** | -5.22*** | 2.84*** |
|  | (1.368) | (1.316) | (2.194) | (1.299) | (1.003) | (0.473) |
| Stay at home requirements |  |  |  |  |  |  |
| *Recommend not leaving house* | 3.40** | 0.93 | 3.91* | 1.89† | 3.57*** | -0.93** |
|  | (1.172) | (0.815) | (1.566) | (1.071) | (0.723) | (0.314) |
| *Require not leaving (loose)* | -6.00*** | -6.71*** | -5.35* | -4.59*** | -5.04*** | 2.98*** |
|  | (1.315) | (1.124) | (2.227) | (1.201) | (0.883) | (0.373) |
| *Require not leaving (strict)* | -19.5*** | -25.7*** | -18.2*** | -15.0*** | -13.0*** | 9.68*** |
|  | (1.757) | (1.526) | (3.059) | (1.809) | (1.349) | (0.569) |
| Internal movement |  |  |  |  |  |  |
| *Recommend movement restriction* | -9.05*** | -6.87*** | -7.09*** | -10.4*** | -3.70*** | 1.98*** |
|  | (0.841) | (0.665) | (1.714) | (0.828) | (0.530) | (0.258) |
| *Restrict movement* | -7.75*** | -6.59*** | -1.99 | -8.96*** | -6.41*** | 1.97*** |
|  | (1.180) | (0.836) | (2.681) | (1.086) | (0.702) | (0.355) |
| % population ages 65 | -0.023 | -0.71*** | -0.35 | -0.30† | -0.36*** | -0.032 |
| and above | (0.135) | (0.141) | (0.396) | (0.166) | (0.0924) | (0.0411) |
| Population density (per sq. km) | -0.013** | -0.0014 | -0.0048 | -0.0051 | -0.010** | 0.0049*** |
| | (0.00419) | (0.00379) | (0.0108) | (0.00590) | (0.00357) | (0.00119) |
| Unemployment (% labour force) | -0.84*** | -0.37*** | -1.70*** | -0.66*** | -0.27*** | 0.21*** |
| | (0.0978) | (0.100) | (0.250) | (0.0995) | (0.0639) | (0.0320) |
| GDP per capita (2010 US$ constant) | 1.74† | 6.72*** | 19.7*** | 3.97** | -0.57 | -1.11** |
| | (1.057) | (1.018) | (3.100) | (1.250) | (0.789) | (0.413) |
| Urban population (% total) | -0.18*** | -0.25*** | -0.99*** | -0.34*** | -0.10** | 0.078*** |
| | (0.0382) | (0.0413) | (0.102) | (0.0531) | (0.0336) | (0.0144) |
| Average temperature (tenths of °C) | 0.030*** | 0.025*** | 0.19*** | 0.037*** | -0.0019 | -0.015*** |
| | (0.00436) | (0.00402) | (0.0103) | (0.00475) | (0.00336) | (0.00138) |
| Average household size | -1.27 | -3.60* | -12.7* | -2.22 | 0.74 | 0.27 |
| | (1.075) | (1.506) | (5.192) | (1.485) | (0.771) | (0.449) |
| Constant | 8.05 | -23.3* | -81.8* | 0.73 | 19.5** | 5.31 |
| | (10.31) | (11.73) | (38.70) | (12.70) | (7.033) | (4.377) |
| Observations | 64800 | 64613 | 58858 | 62958 | 67073 | 58284 |
| Number of clusters | 796 | 785 | 738 | 761 | 798 | 741 |
| Prob. $> \chi^2$ | 0.000 | 0.000 | 0.000 | 0.000 | 0.000 | 0.000 |
| $R^2_{between}$ | 0.849 | 0.578 | 0.367 | 0.821 | 0.782 | 0.823 |
| $R^2_{within}$ | 0.519 | 0.427 | 0.193 | 0.217 | 0.512 | 0.557 |
| $R^2_{overall}$ | 0.789 | 0.538 | 0.290 | 0.700 | 0.743 | 0.771 |

Notes: Results corresponds to Figure 1 in the main text. Random-effects GLS regression estimates. Standard errors (clustered at regional level) in parentheses. † $p < .10$; * $p < .05$; ** $p < .01$; *** $p < .001$. Reference categories are: *Before WHO declares COVID-19 as pandemic*, *Weekdays* and *No measures taken* for all government response indicators.

**Table S2 | Change in visits to six location categories predicted by average individual risk preference before and after pandemic declaration.**

|  | Retail & recreation | Grocery & pharmacy | Parks | Transit stations | Workplaces | Residential |
|---|---|---|---|---|---|---|
| Risk-taking | -1.55 | -4.45*** | -0.46 | -3.53* | -2.40** | -0.067 |
|  | (1.078) | (0.898) | (1.963) | (1.382) | (0.797) | (0.234) |
| Pandemic declaration | -11.3*** | -0.99 | -7.30*** | -12.0*** | -8.10*** | 3.60*** |
|  | (0.879) | (0.705) | (1.379) | (0.833) | (0.642) | (0.285) |
| Pandemic declaration *Risk-taking | 6.72*** | 5.98*** | 11.9*** | 7.17*** | 4.02*** | -0.27 |
|  | (1.166) | (1.013) | (2.449) | (1.422) | (0.871) | (0.423) |
| Weekends | -4.40*** | -3.98*** | -4.55*** | -0.80*** | 8.27*** | -3.28*** |
|  | (0.132) | (0.167) | (0.447) | (0.189) | (0.215) | (0.110) |
| Days after first death | 0.047* | 0.14*** | 0.062† | -0.11*** | 0.048** | 0.017* |
|  | (0.0211) | (0.0167) | (0.0350) | (0.0236) | (0.0187) | (0.00831) |
| ln(# confirmed cases+1) | -2.41*** | -1.94*** | 1.52† | 1.60*** | -1.58*** | 0.14 |
|  | (0.461) | (0.390) | (0.871) | (0.471) | (0.354) | (0.164) |
| School |  |  |  |  |  |  |
| *Recommend closing* | -7.23*** | 1.30 | -30.6*** | 2.46 | 4.20*** | 2.49*** |
|  | (2.142) | (1.099) | (5.076) | (1.632) | (1.047) | (0.583) |
| *Require closing (some)* | 6.78*** | 1.18 | 15.7*** | 5.37* | 0.18 | -1.44*** |
|  | (1.845) | (1.677) | (3.227) | (2.731) | (1.248) | (0.287) |
| *Require closing* | -6.48*** | -5.37*** | -3.95*** | -2.55*** | -5.82*** | 2.48*** |
|  | (0.796) | (0.603) | (1.164) | (0.747) | (0.611) | (0.235) |
| Workplace closing |  |  |  |  |  |  |
| *Recommend closing* | -1.69† | 5.49*** | 8.33*** | -2.95** | 0.93 | -0.44 |
|  | (0.881) | (0.886) | (1.351) | (0.910) | (0.832) | (0.302) |
| *Require closing (some)* | -23.1*** | -6.99*** | -2.45 | -15.5*** | -11.6*** | 4.01*** |
|  | (1.206) | (1.091) | (1.817) | (1.273) | (0.876) | (0.369) |
| *Require closing* | -16.4*** | -4.17*** | -5.07* | -17.1*** | -9.51*** | 4.37*** |
|  | (1.316) | (1.182) | (2.403) | (1.345) | (0.951) | (0.433) |
| Public events |  |  |  |  |  |  |
| *Recommend cancelling* | -0.85 | 2.40* | -5.91*** | -4.61** | -2.45** | 1.88*** |
|  | (1.172) | (1.009) | (1.313) | (1.496) | (0.798) | (0.235) |
| *Require cancelling* | -5.51*** | -0.84 | -4.83** | -4.32*** | -4.36*** | 2.47*** |
|  | (0.711) | (0.686) | (1.637) | (0.863) | (0.678) | (0.230) |
| Restrictions on gatherings |  |  |  |  |  |  |
| *Above 1000 people* | 11.1*** | 9.52*** | 2.61 | 4.90*** | 7.09*** | -3.89*** |
|  | (1.015) | (1.093) | (2.042) | (1.264) | (0.888) | (0.320) |
| *101-1000 people* | 4.70*** | 7.89*** | 1.92 | 1.96 | 5.42*** | -2.39*** |
|  | (1.263) | (1.086) | (2.100) | (1.380) | (0.875) | (0.327) |
| *11-100 people* | -2.10 | -1.75 | -6.58** | -4.46** | -5.11*** | 0.049 |
|  | (1.418) | (1.267) | (2.381) | (1.535) | (1.103) | (0.444) |
| *10 people or less* | -2.12† | -0.95 | -4.27* | -4.00** | -2.92** | -0.19 |
|  | (1.213) | (1.083) | (1.923) | (1.277) | (0.929) | (0.363) |
| Public transport |  |  |  |  |  |  |
| *Recommend closing* | -8.43*** | -9.79*** | -10.1*** | -8.61*** | -10.8*** | 4.30*** |
|  | (0.792) | (0.696) | (1.558) | (0.772) | (0.613) | (0.254) |
| *Require closing* | -7.14*** | -11.3*** | -17.5*** | -6.31*** | -5.13*** | 2.82*** |
|  | (1.359) | (1.317) | (2.194) | (1.295) | (1.000) | (0.475) |
| Stay at home requirements |  |  |  |  |  |  |
| *Recommend not leaving house* | 3.46** | 0.96 | 4.06** | 1.96† | 3.58*** | -0.94** |
|  | (1.180) | (0.831) | (1.571) | (1.097) | (0.744) | (0.317) |
| *Require not leaving (loose)* | -6.22*** | -6.93*** | -5.55* | -4.76*** | -5.17*** | 2.98*** |
|  | (1.310) | (1.128) | (2.221) | (1.209) | (0.890) | (0.372) |
| *Require not leaving (strict)* | -19.7*** | -25.9*** | -18.6*** | -15.3*** | -13.2*** | 9.68*** |
|  | (1.720) | (1.530) | (3.022) | (1.783) | (1.329) | (0.570) |
| Internal movement |  |  |  |  |  |  |
| *Recommend movement restriction* | -8.38*** | -6.28*** | -6.26*** | -9.79*** | -3.33*** | 1.96*** |
|  | (0.843) | (0.675) | (1.724) | (0.836) | (0.533) | (0.255) |
| *Restrict movement* | -7.68*** | -6.53*** | -2.36 | -9.04*** | -6.41*** | 1.98*** |
|  | (1.157) | (0.823) | (2.654) | (1.065) | (0.696) | (0.354) |
| % population ages 65 | -0.067 | -0.75*** | -0.44 | -0.35* | -0.38*** | -0.030 |
| and above | (0.135) | (0.143) | (0.402) | (0.167) | (0.0915) | (0.0413) |
| Population density (per sq. km) | -0.013** | -0.0017 | -0.0056 | -0.0055 | -0.011** | 0.0049*** |
| | (0.00421) | (0.00381) | (0.0108) | (0.00591) | (0.00357) | (0.00120) |
| Unemployment (% labour force) | -0.85*** | -0.38*** | -1.71*** | -0.67*** | -0.28*** | 0.21*** |
| | (0.0981) | (0.102) | (0.251) | (0.101) | (0.0644) | (0.0321) |
| GDP per capita (2010 US$ constant) | 1.96† | 6.91*** | 20.2*** | 4.28*** | -0.45 | -1.12** |
| | (1.065) | (1.028) | (3.129) | (1.260) | (0.787) | (0.412) |
| Urban population (% total) | -0.19*** | -0.25*** | -1.00*** | -0.35*** | -0.11** | 0.079*** |
| | (0.0386) | (0.0415) | (0.102) | (0.0533) | (0.0336) | (0.0144) |
| Average temperature (tenths of °C) | 0.032*** | 0.027*** | 0.19*** | 0.039*** | -0.00050 | -0.015*** |
| | (0.00428) | (0.00401) | (0.0103) | (0.00468) | (0.00332) | (0.00137) |
| Average household size | -1.50 | -3.79* | -13.1* | -2.42 | 0.59 | 0.28 |
| | (1.098) | (1.541) | (5.299) | (1.517) | (0.760) | (0.450) |
| Constant | 7.47 | -23.9* | -84.7* | -0.93 | 19.2** | 5.37 |
| | (10.44) | (11.94) | (39.29) | (12.88) | (7.014) | (4.379) |
| Observations | 64800 | 64613 | 58858 | 62958 | 67073 | 58284 |
| Number of clusters | 796 | 785 | 738 | 761 | 798 | 741 |
| Prob. > $\chi^2$ | 0.000 | 0.000 | 0.000 | 0.000 | 0.000 | 0.000 |
| $R^2_{between}$ | 0.850 | 0.581 | 0.372 | 0.823 | 0.783 | 0.823 |
| $R^2_{within}$ | 0.520 | 0.426 | 0.191 | 0.218 | 0.512 | 0.557 |
| $R^2_{overall}$ | 0.791 | 0.540 | 0.291 | 0.702 | 0.744 | 0.771 |

Notes: Results corresponds to Figure 2 in the main text. Random-effects GLS regression estimates. Standard errors (clustered at regional level) in parentheses. † $p < .10$; * $p < .05$; ** $p < .01$; *** $p < .001$. Reference categories are: *Before WHO declares COVID-19 as pandemic*, *Weekdays* and *No measures taken* for all government response indicators.

**Table S3 | Visitation pattern by weekdays and weekends over average individual risk preference.**

|  | Retail & recreation | Grocery & pharmacy | Parks | Transit stations | Workplaces | Residential |
|---|---|---|---|---|---|---|
| Risk-taking | 2.33* | -0.73 | 7.00** | 1.04 | -0.064 | -0.057 |
|  | (1.184) | (1.050) | (2.537) | (1.352) | (0.862) | (0.391) |
| Weekends | -4.27*** | -3.92*** | -4.39*** | -0.72*** | 8.34*** | -3.35*** |
|  | (0.130) | (0.164) | (0.450) | (0.185) | (0.213) | (0.109) |
| Weekends*Risk-taking | 2.01*** | 0.92* | 2.26* | 1.18* | 1.38** | -0.79** |
|  | (0.318) | (0.411) | (1.015) | (0.502) | (0.506) | (0.260) |
| Pandemic declaration | -11.8*** | -1.42* | -8.38*** | -12.5*** | -8.38*** | 3.62*** |
|  | (0.879) | (0.711) | (1.373) | (0.828) | (0.648) | (0.287) |
| Days after first death | 0.042† | 0.13*** | 0.056 | -0.12*** | 0.046* | 0.017* |
|  | (0.0216) | (0.0171) | (0.0351) | (0.0242) | (0.0190) | (0.00831) |
| ln(# confirmed cases+1) | -2.31*** | -1.86*** | 1.70† | 1.72*** | -1.52*** | 0.13 |
|  | (0.466) | (0.395) | (0.881) | (0.482) | (0.358) | (0.165) |
| School |  |  |  |  |  |  |
| *Recommend closing* | -7.77*** | 0.90 | -31.1*** | 2.07 | 3.84*** | 2.57*** |
|  | (2.111) | (1.086) | (5.010) | (1.617) | (1.033) | (0.585) |
| *Require closing (some)* | 6.62*** | 1.05 | 15.5*** | 5.15† | 0.066 | -1.41*** |
|  | (1.834) | (1.670) | (3.189) | (2.729) | (1.246) | (0.285) |
| *Require closing* | -6.68*** | -5.54*** | -4.29*** | -2.77*** | -5.94*** | 2.50*** |
|  | (0.763) | (0.599) | (1.151) | (0.737) | (0.598) | (0.233) |
| Workplace closing |  |  |  |  |  |  |
| *Recommend closing* | -1.69† | 5.48*** | 8.02*** | -3.09*** | 0.90 | -0.44 |
|  | (0.907) | (0.904) | (1.346) | (0.930) | (0.845) | (0.302) |
| *Require closing (some)* | -23.5*** | -7.33*** | -3.30† | -16.0*** | -11.9*** | 4.03*** |
|  | (1.224) | (1.097) | (1.825) | (1.290) | (0.881) | (0.371) |
| *Require closing* | -16.8*** | -4.49*** | -5.76* | -17.5*** | -9.78*** | 4.38*** |
|  | (1.334) | (1.183) | (2.416) | (1.367) | (0.956) | (0.432) |
| Public events |  |  |  |  |  |  |
| *Recommend cancelling* | -0.64 | 2.58* | -5.57*** | -4.34** | -2.31** | 1.86*** |
|  | (1.180) | (1.017) | (1.323) | (1.503) | (0.791) | (0.233) |
| *Require cancelling* | -5.06*** | -0.42 | -3.91* | -3.79*** | -4.06*** | 2.45*** |
|  | (0.717) | (0.687) | (1.652) | (0.869) | (0.675) | (0.229) |
| Restrictions on gatherings |  |  |  |  |  |  |
| *Above 1000 people* | 10.5*** | 8.96*** | 1.72 | 4.18*** | 6.69*** | -3.87*** |
|  | (1.003) | (1.091) | (2.009) | (1.258) | (0.883) | (0.320) |
| *101-1000 people* | 5.04*** | 8.15*** | 2.64 | 2.34† | 5.63*** | -2.43*** |
|  | (1.246) | (1.078) | (2.041) | (1.367) | (0.862) | (0.323) |
| *11-100 people* | -1.23 | -1.03 | -5.06* | -3.54* | -4.60*** | -0.0035 |
|  | (1.397) | (1.245) | (2.454) | (1.505) | (1.083) | (0.430) |
| *10 people or less* | -1.65 | -0.58 | -3.59† | -3.51** | -2.65** | -0.21 |
|  | (1.191) | (1.065) | (1.877) | (1.257) | (0.914) | (0.357) |
| Public transport |  |  |  |  |  |  |
| *Recommend closing* | -8.53*** | -9.88*** | -10.4*** | -8.78*** | -10.8*** | 4.31*** |
|  | (0.795) | (0.702) | (1.560) | (0.770) | (0.613) | (0.253) |
| *Require closing* | -7.46*** | -11.6*** | -18.3*** | -6.75*** | -5.24*** | 2.84*** |
|  | (1.367) | (1.317) | (2.194) | (1.298) | (1.003) | (0.473) |
| Stay at home requirements |  |  |  |  |  |  |
| *Recommend not leaving house* | 3.41** | 0.94 | 3.92* | 1.90† | 3.58*** | -0.94** |
|  | (1.172) | (0.815) | (1.565) | (1.071) | (0.722) | (0.314) |
| *Require not leaving (loose)* | -5.98*** | -6.70*** | -5.33* | -4.58*** | -5.03*** | 2.97*** |
|  | (1.315) | (1.123) | (2.227) | (1.201) | (0.881) | (0.373) |
| *Require not leaving (strict)* | -19.5*** | -25.7*** | -18.2*** | -15.0*** | -12.9*** | 9.67*** |
|  | (1.756) | (1.526) | (3.060) | (1.808) | (1.348) | (0.569) |
| Internal movement |  |  |  |  |  |  |
| *Recommend movement restriction* | -9.10*** | -6.89*** | -7.14*** | -10.4*** | -3.73*** | 2.00*** |
|  | (0.841) | (0.665) | (1.716) | (0.828) | (0.528) | (0.258) |
| *Restrict movement* | -7.76*** | -6.59*** | -1.99 | -8.96*** | -6.41*** | 1.97*** |
|  | (1.180) | (0.835) | (2.681) | (1.085) | (0.701) | (0.355) |
| % population ages 65 and above | -0.019 | -0.71*** | -0.35 | -0.30† | -0.36*** | -0.033 |
|  | (0.135) | (0.142) | (0.396) | (0.166) | (0.0926) | (0.0412) |
| Population density (per sq. km) | -0.013** | -0.0014 | -0.0049 | -0.0051 | -0.010** | 0.0049*** |
| | (0.00420) | (0.00380) | (0.0108) | (0.00590) | (0.00357) | (0.00119) |
| Unemployment (% labour force) | -0.84*** | -0.37*** | -1.70*** | -0.66*** | -0.27*** | 0.21*** |
| | (0.0976) | (0.100) | (0.249) | (0.0994) | (0.0639) | (0.0320) |
| GDP per capita (2010 US$ constant) | 1.72 | 6.71*** | 19.6*** | 3.96** | -0.57 | -1.10** |
| | (1.056) | (1.018) | (3.101) | (1.251) | (0.790) | (0.413) |
| Urban population (% total) | -0.18*** | -0.25*** | -0.99*** | -0.34*** | -0.10** | 0.078*** |
| | (0.0382) | (0.0413) | (0.102) | (0.0531) | (0.0336) | (0.0144) |
| Average temperature (tenths of °C) | 0.030*** | 0.025*** | 0.19*** | 0.037*** | -0.0020 | -0.015*** |
| | (0.00436) | (0.00402) | (0.0103) | (0.00475) | (0.00336) | (0.00139) |
| Average household size | -1.26 | -3.59* | -12.7* | -2.22 | 0.74 | 0.27 |
| | (1.075) | (1.506) | (5.197) | (1.486) | (0.770) | (0.449) |
| Constant | 8.12 | -23.3* | -81.7* | 0.78 | 19.5** | 5.29 |
| | (10.30) | (11.73) | (38.73) | (12.71) | (7.036) | (4.379) |
| Observations | 64800 | 64613 | 58858 | 62958 | 67073 | 58284 |
| Number of clusters | 796 | 785 | 738 | 761 | 798 | 741 |
| Prob. > $\chi^2$ | 0.000 | 0.000 | 0.000 | 0.000 | 0.000 | 0.000 |
| $R^2_{between}$ | 0.849 | 0.579 | 0.367 | 0.822 | 0.782 | 0.823 |
| $R^2_{within}$ | 0.520 | 0.427 | 0.193 | 0.216 | 0.511 | 0.557 |
| $R^2_{overall}$ | 0.790 | 0.538 | 0.291 | 0.700 | 0.743 | 0.771 |

Notes: Results corresponds to Figure 3 in the main text. Random-effects GLS regression estimates. Standard errors (clustered at regional level) in parentheses. † $p < .10$; * $p < .05$; ** $p < .01$; *** $p < .001$. Reference categories are: *Before WHO declares COVID-19 as pandemic*, *Weekdays* and *No measures taken* for all government response indicators.

**Table S4 | Mediation from risk preference to change in weekends and weekdays visiting pattern pre- and post-pandemic declaration.**

|  | Retail & recreation | Grocery & pharmacy | Parks | Transit stations | Workplaces | Residential |
|---|---|---|---|---|---|---|
| Risk-taking | -1.22 | -3.93*** | -0.11 | -3.01* | -1.94* | -0.072 |
|  | (1.079) | (0.879) | (1.913) | (1.360) | (0.848) | (0.242) |
| Weekends | 1.51*** | 1.22*** | -1.16† | 1.87*** | 1.20*** | -0.43*** |
|  | (0.205) | (0.164) | (0.623) | (0.245) | (0.192) | (0.0542) |
| Weekends*Risk-taking | -1.32* | -1.88*** | -1.48 | -1.88* | -1.28** | 0.10 |
|  | (0.552) | (0.399) | (1.387) | (0.762) | (0.469) | (0.152) |
| Pandemic declaration | -8.93*** | 1.11† | -5.97*** | -11.0*** | -11.0*** | 4.71*** |
|  | (0.861) | (0.675) | (1.336) | (0.813) | (0.634) | (0.285) |
| Pandemic declaration *Risk-taking | 5.46*** | 4.87*** | 10.3*** | 5.94*** | 2.80** | 0.057 |
|  | (1.131) | (0.958) | (2.346) | (1.359) | (0.886) | (0.443) |
| Weekends*Pandemic declaration | -8.52*** | -7.54*** | -4.67*** | -3.75*** | 10.4*** | -4.36*** |
|  | (0.243) | (0.261) | (0.582) | (0.266) | (0.249) | (0.124) |
| Weekends*Pandemic declaration*Risk-taking | 5.04*** | 4.27*** | 5.99*** | 4.70*** | 4.01*** | -1.40*** |
|  | (0.707) | (0.698) | (1.532) | (0.884) | (0.665) | (0.290) |
| Days after first death | 0.049* | 0.14*** | 0.063† | -0.11*** | 0.045* | 0.019* |
|  | (0.0211) | (0.0166) | (0.0349) | (0.0236) | (0.0188) | (0.00833) |
| ln(# confirmed cases+1) | -2.38*** | -1.92*** | 1.55† | 1.62*** | -1.58*** | 0.12 |
|  | (0.459) | (0.388) | (0.871) | (0.470) | (0.356) | (0.165) |
| School |  |  |  |  |  |  |
| *Recommend closing* | -6.88** | 1.63 | -30.5*** | 2.53 | 3.25** | 2.92*** |
|  | (2.129) | (1.105) | (5.095) | (1.632) | (1.040) | (0.604) |
| *Require closing (some)* | 6.85*** | 1.25 | 15.9*** | 5.39* | 0.046 | -1.40*** |
|  | (1.849) | (1.689) | (3.232) | (2.735) | (1.242) | (0.284) |
| *Require closing* | -6.49*** | -5.39*** | -3.95*** | -2.54*** | -5.80*** | 2.50*** |
|  | (0.807) | (0.612) | (1.169) | (0.751) | (0.598) | (0.230) |
| Workplace closing |  |  |  |  |  |  |
| *Recommend closing* | -1.67† | 5.51*** | 8.36*** | -2.94** | 0.98 | -0.46 |
|  | (0.885) | (0.893) | (1.353) | (0.913) | (0.818) | (0.299) |
| *Require closing (some)* | -23.1*** | -7.03*** | -2.43 | -15.5*** | -11.6*** | 3.94*** |
|  | (1.206) | (1.090) | (1.817) | (1.272) | (0.876) | (0.370) |
| *Require closing* | -16.5*** | -4.21*** | -5.07* | -17.1*** | -9.49*** | 4.35*** |
|  | (1.314) | (1.182) | (2.404) | (1.344) | (0.954) | (0.434) |
| Public events |  |  |  |  |  |  |
| *Recommend cancelling* | -1.02 | 2.23* | -5.98*** | -4.69** | -2.15** | 1.70*** |
|  | (1.170) | (1.006) | (1.319) | (1.494) | (0.794) | (0.232) |
| *Require cancelling* | -4.93*** | -0.31 | -4.58** | -4.08*** | -5.09*** | 2.74*** |
|  | (0.708) | (0.689) | (1.632) | (0.860) | (0.687) | (0.237) |
| Restrictions on gatherings |  |  |  |  |  |  |
| *Above 1000 people* | 10.7*** | 9.13*** | 2.40 | 4.72*** | 7.69*** | -4.10*** |
|  | (1.014) | (1.094) | (2.043) | (1.264) | (0.893) | (0.323) |
| *101-1000 people* | 4.20*** | 7.43*** | 1.75 | 1.74 | 6.13*** | -2.64*** |
|  | (1.266) | (1.095) | (2.106) | (1.384) | (0.870) | (0.325) |
| *11-100 people* | -2.82* | -2.39† | -6.97** | -4.79** | -4.18*** | -0.26 |
|  | (1.418) | (1.272) | (2.390) | (1.534) | (1.109) | (0.449) |
| *10 people or less* | -2.67* | -1.44 | -4.56* | -4.26*** | -2.22* | -0.47 |
|  | (1.220) | (1.087) | (1.926) | (1.279) | (0.924) | (0.364) |
| Public transport |  |  |  |  |  |  |
| *Recommend closing* | -8.47*** | -9.83*** | -10.1*** | -8.65*** | -10.8*** | 4.27*** |
|  | (0.789) | (0.696) | (1.561) | (0.771) | (0.616) | (0.256) |
| *Require closing* | -7.22*** | -11.4*** | -17.6*** | -6.36*** | -5.14*** | 2.81*** |
|  | (1.352) | (1.314) | (2.196) | (1.291) | (1.007) | (0.481) |
| Stay at home requirements |  |  |  |  |  |  |
| *Recommend not leaving house* | 3.78** | 1.23 | 4.31** | 2.15* | 3.36*** | -0.79* |
|  | (1.189) | (0.835) | (1.569) | (1.096) | (0.735) | (0.315) |
| *Require not leaving (loose)* | -6.07*** | -6.81*** | -5.40* | -4.65*** | -5.21*** | 3.03*** |
|  | (1.311) | (1.127) | (2.223) | (1.207) | (0.886) | (0.373) |

|  | (1) | (2) | (3) | (4) | (5) | (6) |
|---|---|---|---|---|---|---|
| *Require not leaving (strict)* | -19.5*** | -25.8*** | -18.4*** | -15.2*** | -13.1*** | 9.75*** |
|  | (1.715) | (1.531) | (3.024) | (1.779) | (1.335) | (0.570) |
| Internal movement |  |  |  |  |  |  |
| *Recommend movement restriction* | -8.33*** | -6.23*** | -6.26*** | -9.78*** | -3.50*** | 2.02*** |
|  | (0.837) | (0.674) | (1.725) | (0.833) | (0.534) | (0.256) |
| *Restrict movement* | -7.66*** | -6.49*** | -2.33 | -9.03*** | -6.45*** | 1.98*** |
|  | (1.152) | (0.827) | (2.662) | (1.063) | (0.699) | (0.353) |
| % population ages 65 and above | -0.080 | -0.76*** | -0.45 | -0.36* | -0.37*** | -0.038 |
|  | (0.135) | (0.142) | (0.402) | (0.167) | (0.0913) | (0.0417) |
| Population density (per sq. km) | -0.013** | -0.0016 | -0.0056 | -0.0055 | -0.011** | 0.0050*** |
|  | (0.00420) | (0.00380) | (0.0108) | (0.00590) | (0.00358) | (0.00120) |
| Unemployment (% labour force) | -0.85*** | -0.38*** | -1.71*** | -0.66*** | -0.28*** | 0.22*** |
|  | (0.0977) | (0.102) | (0.251) | (0.101) | (0.0644) | (0.0322) |
| GDP per capita (2010 US$ constant) | 1.81† | 6.80*** | 20.1*** | 4.22*** | -0.29 | -1.15** |
|  | (1.062) | (1.024) | (3.126) | (1.259) | (0.785) | (0.414) |
| Urban population (% total) | -0.18*** | -0.25*** | -1.00*** | -0.35*** | -0.11*** | 0.081*** |
|  | (0.0385) | (0.0413) | (0.102) | (0.0532) | (0.0336) | (0.0145) |
| Average temperature (tenths of °C) | 0.030*** | 0.025*** | 0.19*** | 0.038*** | 0.0020 | -0.016*** |
|  | (0.00427) | (0.00398) | (0.0102) | (0.00468) | (0.00330) | (0.00137) |
| Average household size | -1.47 | -3.76* | -13.1* | -2.41 | 0.54 | 0.29 |
|  | (1.086) | (1.527) | (5.297) | (1.515) | (0.748) | (0.459) |
| Constant | 6.88 | -24.7* | -85.1* | -1.23 | 20.0** | 4.95 |
|  | (10.38) | (11.87) | (39.27) | (12.87) | (6.943) | (4.428) |
| Observations | 64800 | 64613 | 58858 | 62958 | 67073 | 58284 |
| Number of clusters | 796 | 785 | 738 | 761 | 798 | 741 |
| Prob. > $\chi^2$ | 0.000 | 0.000 | 0.000 | 0.000 | 0.000 | 0.000 |
| $R^2_{between}$ | 0.854 | 0.587 | 0.374 | 0.824 | 0.790 | 0.830 |
| $R^2_{within}$ | 0.523 | 0.429 | 0.192 | 0.218 | 0.509 | 0.555 |
| $R^2_{overall}$ | 0.794 | 0.545 | 0.293 | 0.703 | 0.750 | 0.777 |

Notes: Results corresponds to Figure 4 in the main text and Supplementary Figures S1 and S2. Random-effects GLS regression estimates. Standard errors (clustered at regional level) in parentheses. † $p < .10$; * $p < .05$; ** $p < .01$; *** $p < .001$. Reference categories are: *Before WHO declares COVID-19 as pandemic, Weekdays* and *No measures taken* for all government response indicators.

**Table S5 | Change of mobility patterns based on risk preference and share of population.**

|  | Retail & recreation | Grocery & pharmacy | Parks | Transit stations | Workplaces | Residential |
|---|---|---|---|---|---|---|
| Risk-taking | 7.11** | -2.96 | 12.1** | 2.67 | 2.10 | 1.82** |
|  | (2.399) | (2.149) | (4.238) | (3.022) | (1.785) | (0.612) |
| % Population ages 65 and above | -0.23 | -0.60*** | -0.56 | -0.36* | -0.45*** | -0.13** |
|  | (0.148) | (0.164) | (0.484) | (0.153) | (0.0965) | (0.0489) |
| Risk-taking*% Population ages 65 and above | -0.39* | 0.23 | -0.40 | -0.12 | -0.16 | -0.18*** |
|  | (0.169) | (0.159) | (0.371) | (0.199) | (0.115) | (0.0498) |
| Weekends | -4.39*** | -3.97*** | -4.54*** | -0.79*** | 8.28*** | -3.28*** |
|  | (0.132) | (0.167) | (0.446) | (0.189) | (0.215) | (0.110) |
| Pandemic declaration | -11.8*** | -1.42* | -8.39*** | -12.6*** | -8.38*** | 3.62*** |
|  | (0.879) | (0.711) | (1.374) | (0.828) | (0.649) | (0.286) |
| Days after first death | 0.042† | 0.13*** | 0.056 | -0.12*** | 0.046* | 0.017* |
|  | (0.0216) | (0.0172) | (0.0351) | (0.0242) | (0.0190) | (0.00831) |
| $ln$(# confirmed cases+1) | -2.31*** | -1.87*** | 1.70† | 1.72*** | -1.52*** | 0.14 |
|  | (0.466) | (0.395) | (0.880) | (0.481) | (0.358) | (0.165) |
| School |  |  |  |  |  |  |
| *Recommend closing* | -7.61*** | 0.97 | -30.9*** | 2.17 | 3.96*** | 2.50*** |
|  | (2.103) | (1.087) | (4.998) | (1.614) | (1.037) | (0.582) |
| *Require closing (some)* | 6.67*** | 1.08 | 15.6*** | 5.18† | 0.10 | -1.44*** |
|  | (1.834) | (1.672) | (3.185) | (2.729) | (1.243) | (0.284) |
| *Require closing* | -6.65*** | -5.53*** | -4.26*** | -2.75*** | -5.92*** | 2.50*** |
|  | (0.763) | (0.599) | (1.151) | (0.736) | (0.598) | (0.234) |
| Workplace closing |  |  |  |  |  |  |
| *Recommend closing* | -1.70† | 5.48*** | 8.01*** | -3.09*** | 0.89 | -0.44 |
|  | (0.906) | (0.903) | (1.346) | (0.930) | (0.845) | (0.302) |
| *Require closing (some)* | -23.4*** | -7.34*** | -3.28† | -16.0*** | -11.9*** | 4.04*** |
|  | (1.226) | (1.099) | (1.828) | (1.293) | (0.883) | (0.371) |
| *Require closing* | -16.8*** | -4.50*** | -5.76* | -17.5*** | -9.78*** | 4.38*** |
|  | (1.334) | (1.183) | (2.417) | (1.368) | (0.956) | (0.432) |
| Public events |  |  |  |  |  |  |
| *Recommend cancelling* | -0.71 | 2.58* | -5.63*** | -4.37** | -2.35** | 1.85*** |
|  | (1.185) | (1.023) | (1.327) | (1.508) | (0.797) | (0.233) |
| *Require cancelling* | -5.03*** | -0.43 | -3.89* | -3.78*** | -4.05*** | 2.46*** |
|  | (0.715) | (0.685) | (1.652) | (0.868) | (0.674) | (0.229) |
| Restrictions on gatherings |  |  |  |  |  |  |
| *Above 1000 people* | 10.5*** | 8.95*** | 1.68 | 4.17*** | 6.67*** | -3.87*** |
|  | (1.002) | (1.090) | (2.010) | (1.257) | (0.883) | (0.320) |
| *101-1000 people* | 4.97*** | 8.11*** | 2.58 | 2.30† | 5.58*** | -2.40*** |
|  | (1.244) | (1.078) | (2.040) | (1.367) | (0.861) | (0.324) |
| *11-100 people* | -1.30 | -1.05 | -5.13* | -3.58* | -4.65*** | 0.0039 |
|  | (1.397) | (1.242) | (2.457) | (1.504) | (1.083) | (0.430) |
| *10 people or less* | -1.69 | -0.58 | -3.63† | -3.53** | -2.67** | -0.21 |
|  | (1.191) | (1.062) | (1.876) | (1.257) | (0.915) | (0.356) |
| Public transport |  |  |  |  |  |  |
| *Recommend closing* | -8.52*** | -9.87*** | -10.4*** | -8.77*** | -10.8*** | 4.31*** |
|  | (0.795) | (0.701) | (1.560) | (0.770) | (0.613) | (0.253) |
| *Require closing* | -7.42*** | -11.6*** | -18.3*** | -6.73*** | -5.22*** | 2.85*** |
|  | (1.369) | (1.316) | (2.192) | (1.299) | (1.004) | (0.472) |
| Stay at home requirements |  |  |  |  |  |  |
| *Recommend not leaving house* | 3.40** | 0.94 | 3.90* | 1.89† | 3.57*** | -0.94** |
|  | (1.173) | (0.815) | (1.567) | (1.072) | (0.723) | (0.314) |
| *Require not leaving (loose)* | -6.00*** | -6.71*** | -5.36* | -4.59*** | -5.05*** | 2.98*** |
|  | (1.315) | (1.123) | (2.228) | (1.201) | (0.883) | (0.372) |
| *Require not leaving (strict)* | -19.5*** | -25.7*** | -18.2*** | -15.0*** | -13.0*** | 9.68*** |
|  | (1.756) | (1.527) | (3.058) | (1.808) | (1.349) | (0.569) |
| Internal movement |  |  |  |  |  |  |
| *Recommend movement restriction* | -9.09*** | -6.84*** | -7.12*** | -10.4*** | -3.72*** | 1.96*** |
|  | (0.839) | (0.663) | (1.711) | (0.828) | (0.530) | (0.259) |
| *Restrict movement* | -7.77*** | -6.57*** | -2.00 | -8.96*** | -6.42*** | 1.96*** |
|  | (1.180) | (0.835) | (2.679) | (1.086) | (0.701) | (0.354) |
| Population density (per sq. km) | -0.012** | -0.0021 | -0.0035 | -0.0047 | -0.0098** | 0.0055*** |
|  | (0.00406) | (0.00362) | (0.0107) | (0.00555) | (0.00342) | (0.00117) |
| Unemployment (% labour | -0.89*** | -0.35** | -1.75*** | -0.67*** | -0.29*** | 0.19*** |
|  | (0.0954) | (0.106) | (0.247) | (0.102) | (0.0663) | (0.0328) |
| force) | | | | | | |
| GDP per capita (2010 US$ constant) | 2.47* | 6.31*** | 20.4*** | 4.20*** | -0.25 | -0.81* |
| | (1.092) | (1.027) | (3.114) | (1.204) | (0.780) | (0.413) |
| Urban population (% total) | -0.18*** | -0.25*** | -0.99*** | -0.34*** | -0.10** | 0.077*** |
| | (0.0382) | (0.0409) | (0.102) | (0.0524) | (0.0335) | (0.0142) |
| Average temperature (tenths of °C) | 0.030*** | 0.025*** | 0.19*** | 0.037*** | -0.0020 | -0.015*** |
| | (0.00434) | (0.00399) | (0.0103) | (0.00474) | (0.00336) | (0.00139) |
| Average household size | -1.69 | -3.34* | -13.2* | -2.32 | 0.57 | 0.024 |
| | (1.158) | (1.525) | (5.422) | (1.559) | (0.785) | (0.423) |
| Constant | 4.96 | -21.7† | -84.0* | -0.26 | 18.1** | 4.59 |
| | (10.47) | (11.44) | (38.82) | (12.52) | (6.918) | (4.184) |
| Observations | 64800 | 64613 | 58858 | 62958 | 67073 | 58284 |
| Number of clusters | 796 | 785 | 738 | 761 | 798 | 741 |
| Prob. > $\chi^2$ | 0.000 | 0.000 | 0.000 | 0.000 | 0.000 | 0.000 |
| $R^2_{between}$ | 0.849 | 0.578 | 0.367 | 0.821 | 0.782 | 0.823 |
| $R^2_{within}$ | 0.522 | 0.428 | 0.196 | 0.217 | 0.512 | 0.566 |
| $R^2_{overall}$ | 0.790 | 0.539 | 0.292 | 0.700 | 0.743 | 0.773 |

Notes: Results corresponds to Figure 5 in the main text. Random-effects GLS regression estimates. Standard errors (clustered at regional level) in parentheses. † $p < .10$; * $p < .05$; ** $p < .01$; *** $p < .001$. Reference categories are: *Before WHO declares COVID-19 as pandemic, Weekdays* and *No measures taken* for all government response indicators.

**Table S6 | Country data availability for mobility (Google), risk preferences (GPS), and government response (OxCGRT)**

| Country | Google | GPS | OxCGRT | Country | Google | GPS | OxCGRT |
|---|---|---|---|---|---|---|---|
| Afghanistan | Yes | Yes | Yes | Lebanon | Yes | No | Yes |
| Algeria | No | Yes | Yes | Lesotho | No | No | Yes |
| Andorra | No | No | Yes | Libya | Yes | No | Yes |
| Angola | Yes | No | Yes | Liechtenstein | Yes | No | No |
| Antigua and Barbuda | Yes | No | No | Lithuania | Yes | Yes | No |
| Argentina | Yes | Yes | Yes | Luxembourg | Yes | No | Yes |
| Aruba | Yes | No | Yes | Madagascar | No | No | Yes |
| Australia | Yes | Yes | Yes | Malawi | No | Yes | Yes |
| Austria | Yes | Yes | Yes | Malaysia | Yes | No | Yes |
| Azerbaijan | No | No | Yes | Mali | Yes | No | Yes |
| Bahrain | Yes | No | Yes | Malta | Yes | No | No |
| Bangladesh | Yes | Yes | Yes | Mauritania | No | No | Yes |
| Barbados | Yes | No | Yes | Mauritius | Yes | No | Yes |
| Belarus | Yes | No | No | Mexico | Yes | Yes | Yes |
| Belgium | Yes | No | Yes | Moldova | Yes | Yes | Yes |
| Belize | Yes | No | Yes | Mongolia | Yes | No | Yes |
| Benin | Yes | No | No | Morocco | No | Yes | No |
| Bermuda | No | No | Yes | Mozambique | Yes | No | Yes |
| Bolivia | Yes | Yes | Yes | Myanmar (Burma) | Yes | No | Yes |
| Bosnia and Herzegovina | Yes | Yes | Yes | Namibia | Yes | No | Yes |
| Botswana | Yes | Yes | Yes | Nepal | Yes | No | No |
| Brazil | Yes | Yes | Yes | Netherlands | Yes | Yes | Yes |
| Brunei | No | No | Yes | New Zealand | Yes | No | Yes |
| Bulgaria | Yes | No | Yes | Nicaragua | Yes | Yes | Yes |
| Burkina Faso | Yes | No | Yes | Niger | Yes | No | Yes |
| Burundi | No | No | Yes | Nigeria | Yes | Yes | Yes |
| Cambodia | Yes | Yes | No | North Macedonia | Yes | No | No |
| Cameroon | Yes | Yes | Yes | Norway | Yes | No | Yes |
| Canada | Yes | Yes | Yes | Oman | Yes | No | Yes |
| Cape Verde | Yes | No | Yes | Pakistan | Yes | Yes | Yes |
| Chad | No | No | Yes | Palestine | No | No | Yes |
| Chile | Yes | Yes | Yes | Panama | Yes | No | Yes |
| China | No | Yes | Yes | Papua New Guinea | Yes | No | Yes |
| Colombia | Yes | Yes | Yes | Paraguay | Yes | No | Yes |
| Costa Rica | Yes | Yes | Yes | Peru | Yes | Yes | Yes |
| Croatia | Yes | Yes | Yes | Philippines | Yes | Yes | Yes |
| Cuba | No | No | Yes | Poland | Yes | Yes | Yes |
| Cyprus | No | No | Yes | Portugal | Yes | Yes | Yes |
| Czechia | Yes | Yes | Yes | Puerto Rico | Yes | No | Yes |
| Côte d'Ivoire | Yes | No | No | Qatar | Yes | No | Yes |
| Democratic Republic of Congo | No | No | Yes | Romania | Yes | Yes | Yes |
| Denmark | Yes | No | Yes | Russia | No | Yes | No |
| Djibouti | No | No | Yes | Rwanda | Yes | Yes | Yes |
| Dominican Republic | Yes | No | Yes | Réunion | Yes | No | No |
| Ecuador | Yes | No | Yes | San Marino | No | No | Yes |
| Egypt | Yes | Yes | Yes | Saudi Arabia | Yes | Yes | Yes |
| El Salvador | Yes | No | Yes | Senegal | Yes | No | No |
| Estonia | Yes | Yes | Yes | Serbia | No | Yes | No |
| Eswatini | No | No | Yes | Seychelles | No | No | Yes |
| Fiji | Yes | No | No | Sierra Leone | No | No | Yes |
| Finland | Yes | Yes | Yes | Singapore | Yes | No | Yes |
| France | Yes | Yes | Yes | Slovakia | Yes | No | Yes |
| Gabon | Yes | No | Yes | Slovenia | Yes | No | Yes |
| Gambia | No | No | Yes | South Africa | Yes | Yes | Yes |
| Georgia | Yes | Yes | No | South Korea | Yes | Yes | Yes |
| Germany | Yes | Yes | Yes | South Sudan | No | No | Yes |
| Ghana | Yes | Yes | Yes | Spain | Yes | Yes | Yes |
| Greece | Yes | Yes | Yes | Sri Lanka | Yes | Yes | Yes |
| Greenland | No | No | Yes | Sudan | No | No | Yes |
| Guam | No | No | Yes | Suriname | No | Yes | No |
| Guatemala | Yes | Yes | Yes | Sweden | Yes | Yes | Yes |
| Guinea-Bissau | Yes | No | No | Switzerland | Yes | Yes | Yes |
| Guyana | No | No | Yes | Syria | No | No | Yes |
| Haiti | Yes | Yes | No | Taiwan | Yes | No | Yes |
| Honduras | Yes | No | Yes | Tajikistan | Yes | No | No |
| Hong Kong | Yes | No | Yes | Tanzania | Yes | Yes | Yes |
| Hungary | Yes | Yes | Yes | Thailand | Yes | Yes | Yes |
| India | Yes | Yes | Yes | The Bahamas | Yes | No | No |
| Indonesia | Yes | Yes | Yes | Togo | Yes | No | No |
| Iran | No | Yes | Yes | Trinidad and Tobago | Yes | No | Yes |
| Iraq | Yes | Yes | Yes | Tunisia | No | No | Yes |
| Ireland | Yes | No | Yes | Turkey | Yes | Yes | Yes |
| Israel | Yes | Yes | Yes | Uganda | Yes | Yes | Yes |
| Italy | Yes | Yes | Yes | Ukraine | No | Yes | No |
| Jamaica | Yes | No | Yes | United Arab Emirates | Yes | Yes | Yes |
| Japan | Yes | Yes | Yes | United Kingdom | Yes | Yes | Yes |
| Jordan | Yes | Yes | Yes | United States | Yes | Yes | Yes |
| Kazakhstan | Yes | Yes | Yes | Uruguay | Yes | No | Yes |
| Kenya | Yes | Yes | Yes | Uzbekistan | No | No | Yes |
| Kosovo | No | No | Yes | Venezuela | Yes | Yes | Yes |
| Kuwait | Yes | No | Yes | Vietnam | Yes | Yes | Yes |
| Kyrgyzstan | Yes | No | Yes | Yemen | Yes | No | No |
| Laos | Yes | No | Yes | Zambia | Yes | No | Yes |
| Latvia | Yes | No | No | Zimbabwe | Yes | Yes | Yes |

Note: GPS = Global Preference Survey. OxCGRT = Oxford COVID-19 Government Response Tracker. Numbers in bracket show the number of regions in the corresponding dataset.

**Table S7 | Robustness checks on overall risk-mobility relationship.**

| Robust 1 | Retail & recreation | Grocery & pharmacy | Parks | Transit stations | Workplaces | Residential |
|---|---|---|---|---|---|---|
| Risk-taking | 2.22† | -0.33 | 8.02** | 1.31 | 0.76 | 0.089 |
|  | (1.183) | (1.006) | (2.535) | (1.352) | (0.854) | (0.413) |
| Pandemic declaration | -12.9*** | -2.48*** | -10.1*** | -13.0*** | -8.93*** | 4.36*** |
|  | (0.911) | (0.744) | (1.379) | (0.822) | (0.645) | (0.323) |
| Weekends | -4.88*** | -4.23*** | -4.95*** | -0.70*** | 8.33*** | -3.39*** |
|  | (0.127) | (0.170) | (0.478) | (0.198) | (0.218) | (0.129) |
| % Population 65+ | -0.18 | -0.94*** | -1.50*** | -0.30† | -0.17† | 0.20*** |
|  | (0.147) | (0.128) | (0.295) | (0.156) | (0.0922) | (0.0489) |
| Controls | Yes | Yes | Yes | Yes | Yes | Yes |
| Observations | 58340 | 57571 | 51715 | 57764 | 63041 | 44546 |
| Number of clusters | 688 | 678 | 609 | 680 | 744 | 524 |
| Prob. $> \chi^2$ | 0.000 | 0.000 | 0.000 | 0.000 | 0.000 | 0.000 |
| $R^2$-between | 0.861 | 0.600 | 0.403 | 0.827 | 0.788 | 0.806 |
| $R^2$-within | 0.415 | 0.431 | 0.339 | 0.168 | 0.513 | 0.489 |
| $R^2$-overall | 0.798 | 0.558 | 0.371 | 0.704 | 0.750 | 0.761 |
| **Robust 2** | Retail & recreation | Grocery & pharmacy | Parks | Transit stations | Workplaces | Residential |
| Risk-taking | 2.55* | -0.28 | 9.46*** | 1.86 | -0.25 | 0.25 |
|  | (1.074) | (0.905) | (2.589) | (1.186) | (0.773) | (0.427) |
| Pandemic declaration | -14.0*** | -4.68*** | -10.9*** | -14.4*** | -10.7*** | 4.33*** |
|  | (1.031) | (0.822) | (1.519) | (0.908) | (0.716) | (0.330) |
| Weekends | -4.73*** | -3.84*** | -4.41*** | -1.42*** | 7.54*** | -3.33*** |
|  | (0.146) | (0.208) | (0.549) | (0.235) | (0.271) | (0.134) |
| % Population 65+ | -0.46** | -0.93*** | -1.83*** | -0.25† | -0.22* | 0.22*** |
|  | (0.155) | (0.128) | (0.319) | (0.143) | (0.103) | (0.0514) |
| Controls | Yes | Yes | Yes | Yes | Yes | Yes |
| Observations | 41146 | 41146 | 41146 | 41146 | 41146 | 41146 |
| Number of clusters | 484 | 484 | 484 | 484 | 484 | 484 |
| Prob. $> \chi^2$ | 0.000 | 0.000 | 0.000 | 0.000 | 0.000 | 0.000 |
| $R^2$-between | 0.849 | 0.570 | 0.391 | 0.840 | 0.771 | 0.807 |
| $R^2$-within | 0.583 | 0.572 | 0.372 | 0.321 | 0.638 | 0.508 |
| $R^2$-overall | 0.803 | 0.571 | 0.377 | 0.773 | 0.752 | 0.763 |
| **Robust 3** | Retail & recreation | Grocery & pharmacy | Parks | Transit stations | Workplaces | Residential |
| Risk-taking | 3.72** | -0.25 | 6.07* | 1.60 | 0.58 | -0.48 |
|  | (1.225) | (1.138) | (2.455) | (1.597) | (0.940) | (0.374) |
| Pandemic declaration | -17.2*** | -3.59*** | -10.7*** | -18.2*** | -13.7*** | 5.91*** |
|  | (0.790) | (0.518) | (1.084) | (0.825) | (0.612) | (0.264) |
| Weekends | -4.44*** | -4.08*** | -4.60*** | -0.87*** | 8.31*** | -3.29*** |
|  | (0.140) | (0.172) | (0.459) | (0.191) | (0.209) | (0.109) |
| % Population 65+ | 0.30* | -0.56*** | -0.0023 | 0.099 | -0.0078 | -0.16** |
|  | (0.146) | (0.160) | (0.396) | (0.192) | (0.102) | (0.0510) |
| Controls | Yes | Yes | Yes | Yes | Yes | Yes |
| Observations | 64800 | 64613 | 58858 | 62958 | 67073 | 58284 |
| Number of clusters | 796 | 785 | 738 | 761 | 798 | 741 |
| Prob. $> \chi^2$ | 0.000 | 0.000 | 0.000 | 0.000 | 0.000 | 0.000 |
| $R^2$-between | 0.846 | 0.568 | 0.386 | 0.808 | 0.768 | 0.802 |
| $R^2$-within | 0.632 | 0.471 | 0.284 | 0.272 | 0.528 | 0.626 |
| $R^2$-overall | 0.809 | 0.537 | 0.344 | 0.697 | 0.730 | 0.764 |

Notes: **Robust 1** = regions with at least one censored values on the outcome mobility measures excluded. **Robust 2** = regions with at least one censored values on any mobility measures excluded. **Robust 3** = government response indicators recoded as no measures taken if policy is not applied countrywide. Random-effects GLS regression estimates. Standard errors (clustered at regional level) in parentheses. † $p < .10$; * $p < .05$; ** $p < .01$; *** $p < .001$. We controlled for number of confirmed cases (in logs), population density, and the set of government response indicators in each regression. Reference categories are: *Before WHO declares COVID-19 as pandemic*, *Weekdays* and *No measures taken*.

**Table S8 | Robustness checks on moderation effect of pandemic declaration on risk-mobility relationship.**

| **Robust 1** | Retail & recreation | Grocery & pharmacy | Parks | Transit stations | Workplaces | Residential |
|---|---|---|---|---|---|---|
| Risk-taking | -1.13 | -4.54*** | 0.91 | -3.33* | -1.78* | 0.24 |
| | (1.016) | (0.806) | (1.885) | (1.340) | (0.794) | (0.223) |
| Pandemic declaration | -12.5*** | -1.92** | -9.06*** | -12.4*** | -8.65*** | 4.33*** |
| | (0.920) | (0.742) | (1.396) | (0.833) | (0.641) | (0.324) |
| Pandemic declaration*Risk-taking | 4.96*** | 6.26*** | 10.5*** | 6.92*** | 3.77*** | -0.22 |
| | (1.210) | (1.075) | (2.477) | (1.514) | (0.880) | (0.515) |
| Controls | Yes | Yes | Yes | Yes | Yes | Yes |
| Observations | 58340 | 57571 | 51715 | 57764 | 63041 | 44546 |
| Number of clusters | 688 | 678 | 609 | 680 | 744 | 524 |
| Prob. $> \chi^2$ | 0.000 | 0.000 | 0.000 | 0.000 | 0.000 | 0.000 |
| $R^2$-between | 0.862 | 0.603 | 0.407 | 0.828 | 0.789 | 0.806 |
| $R^2$-within | 0.418 | 0.432 | 0.337 | 0.170 | 0.512 | 0.489 |
| $R^2$-overall | 0.799 | 0.560 | 0.372 | 0.705 | 0.751 | 0.761 |
| **Robust 2** | Retail & recreation | Grocery & pharmacy | Parks | Transit stations | Workplaces | Residential |
| Risk-taking | -0.65 | -3.10*** | 1.54 | -1.45 | -1.27* | 0.31 |
| | (0.787) | (0.627) | (2.131) | (1.055) | (0.638) | (0.228) |
| Pandemic declaration | -13.4*** | -4.16*** | -9.42*** | -13.8*** | -10.5*** | 4.32*** |
| | (1.053) | (0.835) | (1.564) | (0.937) | (0.729) | (0.331) |
| Pandemic declaration*Risk-taking | 4.70*** | 4.15*** | 11.7*** | 4.88** | 1.50 | -0.092 |
| | (1.388) | (1.190) | (2.695) | (1.698) | (0.995) | (0.548) |
| Controls | Yes | Yes | Yes | Yes | Yes | Yes |
| Observations | 41146 | 41146 | 41146 | 41146 | 41146 | 41146 |
| Number of clusters | 484 | 484 | 484 | 484 | 484 | 484 |
| Prob. $> \chi^2$ | 0.000 | 0.000 | 0.000 | 0.000 | 0.000 | 0.000 |
| $R^2$-between | 0.850 | 0.572 | 0.395 | 0.841 | 0.771 | 0.807 |
| $R^2$-within | 0.587 | 0.574 | 0.371 | 0.326 | 0.639 | 0.508 |
| $R^2$-overall | 0.804 | 0.572 | 0.379 | 0.774 | 0.753 | 0.763 |
| **Robust 3** | Retail & recreation | Grocery & pharmacy | Parks | Transit stations | Workplaces | Residential |
| Risk-taking | 1.78 | -2.17* | 1.30 | -0.38 | 0.39 | -1.17*** |
| | (1.096) | (0.945) | (2.028) | (1.517) | (0.823) | (0.263) |
| Pandemic declaration | -17.2*** | -3.51*** | -10.4*** | -18.1*** | -13.7*** | 5.95*** |
| | (0.795) | (0.527) | (1.087) | (0.831) | (0.615) | (0.261) |
| Pandemic declaration*Risk-taking | 2.98** | 2.93** | 7.07** | 2.93* | 0.28 | 1.07* |
| | (1.083) | (1.025) | (2.382) | (1.322) | (0.923) | (0.443) |
| Controls | Yes | Yes | Yes | Yes | Yes | Yes |
| Observations | 64800 | 64613 | 58858 | 62958 | 67073 | 58284 |
| Number of clusters | 796 | 785 | 738 | 761 | 798 | 741 |
| Prob. $> \chi^2$ | 0.000 | 0.000 | 0.000 | 0.000 | 0.000 | 0.000 |
| $R^2$-between | 0.846 | 0.569 | 0.388 | 0.809 | 0.768 | 0.803 |
| $R^2$-within | 0.631 | 0.470 | 0.282 | 0.271 | 0.528 | 0.629 |
| $R^2$-overall | 0.809 | 0.537 | 0.344 | 0.697 | 0.730 | 0.765 |

Notes: **Robust 1** = regions with at least one censored values on the outcome mobility measures excluded. **Robust 2** = regions with at least one censored values on any mobility measures excluded. **Robust 3** = government response indicators recoded as no measures taken if policy is not applied countrywide. Random-effects GLS regression estimates. Standard errors (clustered at regional level) in parentheses. † $p < .10$; * $p < .05$; ** $p < .01$; *** $p < .001$. We controlled for weekend dummy, share of population over 65, day since first confirmed death, number of confirmed cases (in logs), population density, and the set of government response indicators in each regression. Reference categories are: *Before WHO declares COVID-19 as pandemic*, *Weekdays* and *No measures taken*.

**Table S9 | Robustness checks on weekends-weekdays mobility change with mediation from risk attitude.**

| **Robust 1** | Retail & recreation | Grocery & pharmacy | Parks | Transit stations | Workplaces | Residential |
|---|---|---|---|---|---|---|
| Risk-taking | 1.78 | -0.51 | 7.10** | 0.97 | 0.40 | 0.39 |
|  | (1.190) | (0.995) | (2.484) | (1.360) | (0.873) | (0.445) |
| Weekends | -4.76*** | -4.18*** | -4.69*** | -0.63** | 8.40*** | -3.49*** |
|  | (0.125) | (0.166) | (0.487) | (0.193) | (0.216) | (0.128) |
| Weekends*Risk-taking | 1.62*** | 0.70† | 3.44** | 1.25* | 1.32* | -1.14*** |
|  | (0.308) | (0.406) | (1.092) | (0.541) | (0.511) | (0.305) |
| Controls | Yes | Yes | Yes | Yes | Yes | Yes |
| Observations | 58340 | 57571 | 51715 | 57764 | 63041 | 44546 |
| Number of clusters | 688 | 678 | 609 | 680 | 744 | 524 |
| Prob. $> \chi^2$ | 0.000 | 0.000 | 0.000 | 0.000 | 0.000 | 0.000 |
| $R^2$-between | 0.861 | 0.600 | 0.403 | 0.827 | 0.788 | 0.807 |
| $R^2$-within | 0.416 | 0.431 | 0.339 | 0.168 | 0.513 | 0.489 |
| $R^2$-overall | 0.799 | 0.558 | 0.371 | 0.704 | 0.750 | 0.761 |
| **Robust 2** | Retail & recreation | Grocery & pharmacy | Parks | Transit stations | Workplaces | Residential |
| Risk-taking | 2.02† | -0.49 | 8.30** | 1.35 | -0.78 | 0.58 |
|  | (1.082) | (0.894) | (2.541) | (1.210) | (0.815) | (0.463) |
| Weekends | -4.55*** | -3.77*** | -4.01*** | -1.25*** | 7.73*** | -3.44*** |
|  | (0.140) | (0.208) | (0.563) | (0.236) | (0.276) | (0.133) |
| Weekends*Risk-taking | 2.04*** | 0.80† | 4.47*** | 1.96*** | 2.06** | -1.25*** |
|  | (0.349) | (0.454) | (1.306) | (0.585) | (0.635) | (0.321) |
| Controls | Yes | Yes | Yes | Yes | Yes | Yes |
| Observations | 41146 | 41146 | 41146 | 41146 | 41146 | 41146 |
| Number of clusters | 484 | 484 | 484 | 484 | 484 | 484 |
| Prob. $> \chi^2$ | 0.000 | 0.000 | 0.000 | 0.000 | 0.000 | 0.000 |
| $R^2$-between | 0.849 | 0.571 | 0.391 | 0.840 | 0.771 | 0.807 |
| $R^2$-within | 0.583 | 0.572 | 0.373 | 0.322 | 0.638 | 0.508 |
| $R^2$-overall | 0.803 | 0.571 | 0.378 | 0.773 | 0.753 | 0.764 |
| **Robust 3** | Retail & recreation | Grocery & pharmacy | Parks | Transit stations | Workplaces | Residential |
| Risk-taking | 3.33** | -0.37 | 5.48* | 1.36 | 0.28 | -0.31 |
|  | (1.241) | (1.148) | (2.439) | (1.604) | (0.965) | (0.386) |
| Weekends | -4.36*** | -4.05*** | -4.47*** | -0.82*** | 8.36*** | -3.34*** |
|  | (0.139) | (0.170) | (0.464) | (0.186) | (0.206) | (0.108) |
| Weekends*Risk-taking | 1.46*** | 0.47 | 2.00† | 0.91† | 1.12* | -0.72** |
|  | (0.331) | (0.420) | (1.026) | (0.494) | (0.494) | (0.259) |
| Controls | Yes | Yes | Yes | Yes | Yes | Yes |
| Observations | 64800 | 64613 | 58858 | 62958 | 67073 | 58284 |
| Number of clusters | 796 | 785 | 738 | 761 | 798 | 741 |
| Prob. $> \chi^2$ | 0.000 | 0.000 | 0.000 | 0.000 | 0.000 | 0.000 |
| $R^2$-between | 0.846 | 0.568 | 0.386 | 0.808 | 0.768 | 0.802 |
| $R^2$-within | 0.633 | 0.471 | 0.284 | 0.272 | 0.528 | 0.626 |
| $R^2$-overall | 0.809 | 0.537 | 0.344 | 0.697 | 0.730 | 0.765 |

Notes: **Robust 1** = regions with at least one censored values on the outcome mobility measures excluded. **Robust 2** = regions with at least one censored values on any mobility measures excluded. **Robust 3** = government response indicators recoded as no measures taken if policy is not applied countrywide. Random-effects GLS regression estimates. Standard errors (clustered at regional level) in parentheses. † $p < .10$; * $p < .05$; ** $p < .01$; *** $p < .001$. We controlled for pandemic declaration dummy, day since first confirmed death, share of population over 65, number of confirmed cases (in logs), population density, and the set of government response indicators in each regression. Reference categories are: *Before WHO declares COVID-19 as pandemic*, *Weekdays* and *No measures taken*.

**Table S10 | Robustness checks on the moderating effect of pandemic declaration on weekends-weekdays mobility change based on risk preference.**

| Robust 1 | Retail & recreation | Grocery & pharmacy | Parks | Transit stations | Workplaces | Residential |
|---|---|---|---|---|---|---|
| Risk-taking | -0.88 | -4.05*** | 0.79 | -2.72* | -1.24 | 0.16 |
| | (1.020) | (0.777) | (1.836) | (1.316) | (0.846) | (0.229) |
| Weekends | 1.39*** | 1.06*** | -0.29 | 1.98*** | 1.04*** | -0.38*** |
| | (0.214) | (0.167) | (0.673) | (0.248) | (0.191) | (0.0569) |
| Weekends*Risk-taking | -1.06† | -1.83*** | 0.31 | -2.18** | -1.50** | 0.10 |
| | (0.588) | (0.426) | (1.384) | (0.792) | (0.470) | (0.165) |
| Pandemic declaration | -10.0*** | 0.21 | -7.31*** | -11.4*** | -11.7*** | 5.55*** |
| | (0.905) | (0.707) | (1.338) | (0.813) | (0.631) | (0.327) |
| Pandemic declaration*Risk-taking | 3.95*** | 5.27*** | 9.25*** | 5.53*** | 2.47** | 0.31 |
| | (1.195) | (1.011) | (2.376) | (1.448) | (0.893) | (0.550) |
| Weekends*Pandemic declaration | -8.93*** | -7.59*** | -6.37*** | -3.78*** | 10.7*** | -4.52*** |
| | (0.242) | (0.271) | (0.588) | (0.273) | (0.248) | (0.140) |
| Weekends*Pandemic declaration*Risk-taking | 4.12*** | 3.93*** | 5.03*** | 5.28*** | 4.25*** | -1.82*** |
| | (0.690) | (0.717) | (1.419) | (0.933) | (0.658) | (0.335) |
| Controls | Yes | Yes | Yes | Yes | Yes | Yes |
| Observations | 58340 | 57571 | 51715 | 57764 | 63041 | 44546 |
| Number of clusters | 688 | 678 | 609 | 680 | 744 | 524 |
| Prob. > $\chi^2$ | 0.000 | 0.000 | 0.000 | 0.000 | 0.000 | 0.000 |
| $R^2$-between | 0.866 | 0.609 | 0.409 | 0.829 | 0.796 | 0.815 |
| $R^2$-within | 0.422 | 0.435 | 0.339 | 0.171 | 0.508 | 0.485 |
| $R^2$-overall | 0.803 | 0.566 | 0.374 | 0.707 | 0.757 | 0.767 |
| **Robust 2** | Retail & recreation | Grocery & pharmacy | Parks | Transit stations | Workplaces | Residential |
| Risk-taking | -1.01 | -3.11*** | 0.78 | -1.36 | -0.70 | 0.25 |
| | (0.777) | (0.606) | (2.065) | (1.025) | (0.712) | (0.235) |
| Weekends | 1.79*** | 1.67*** | 0.43 | 2.02*** | 0.63** | -0.41*** |
| | (0.234) | (0.190) | (0.779) | (0.293) | (0.241) | (0.0590) |
| Weekends*Risk-taking | 1.02† | -0.19 | 2.56 | -0.45 | -1.64** | 0.061 |
| | (0.596) | (0.458) | (1.673) | (0.800) | (0.564) | (0.175) |
| Pandemic declaration | -10.9*** | -2.03** | -7.68*** | -12.5*** | -13.3*** | 5.50*** |
| | (1.030) | (0.786) | (1.500) | (0.913) | (0.719) | (0.334) |
| Pandemic declaration*Risk-taking | 4.43** | 3.83*** | 11.0*** | 3.99* | -0.071 | 0.46 |
| | (1.359) | (1.118) | (2.594) | (1.643) | (1.027) | (0.584) |
| Weekends*Pandemic declaration | -9.18*** | -7.87*** | -6.39*** | -4.71*** | 10.3*** | -4.40*** |
| | (0.271) | (0.342) | (0.699) | (0.302) | (0.314) | (0.144) |
| Weekends*Pandemic declaration*Risk-taking | 1.69* | 1.63* | 3.31† | 3.73*** | 5.46*** | -1.92*** |
| | (0.723) | (0.775) | (1.700) | (0.869) | (0.799) | (0.343) |
| Controls | Yes | Yes | Yes | Yes | Yes | Yes |
| Observations | 41146 | 41146 | 41146 | 41146 | 41146 | 41146 |
| Number of clusters | 484 | 484 | 484 | 484 | 484 | 484 |
| Prob. > $\chi^2$ | 0.000 | 0.000 | 0.000 | 0.000 | 0.000 | 0.000 |
| $R^2$-between | 0.854 | 0.578 | 0.398 | 0.842 | 0.778 | 0.814 |
| $R^2$-within | 0.590 | 0.576 | 0.373 | 0.328 | 0.637 | 0.504 |
| $R^2$-overall | 0.808 | 0.578 | 0.381 | 0.776 | 0.758 | 0.769 |
| **Robust 3** | Retail & recreation | Grocery & pharmacy | Parks | Transit stations | Workplaces | Residential |
| Risk-taking | 2.07† | -1.68† | 1.54 | 0.088 | 0.90 | -1.14*** |
| | (1.092) | (0.942) | (1.976) | (1.503) | (0.871) | (0.260) |
| Weekends | 1.64*** | 1.28*** | -1.07† | 2.01*** | 1.36*** | -0.55*** |
| | (0.205) | (0.163) | (0.616) | (0.231) | (0.190) | (0.0564) |
| Weekends*Risk-taking | -1.28* | -1.85*** | -1.21 | -1.73* | -1.39** | -0.036 |
| | (0.546) | (0.406) | (1.380) | (0.719) | (0.460) | (0.147) |
| Pandemic declaration | -14.5*** | -1.15* | -8.90*** | -16.8*** | -16.7*** | 7.10*** |
| | (0.770) | (0.518) | (1.065) | (0.823) | (0.638) | (0.270) |
| Pandemic declaration*Risk-taking | 1.94† | 1.98* | 5.75* | 1.88 | -0.75 | 1.26** |
| | (1.051) | (0.992) | (2.289) | (1.265) | (0.956) | (0.459) |
| Weekends*Pandemic declaration | -8.83*** | -7.82*** | -4.88*** | -4.08*** | 10.2*** | -4.17*** |
| | (0.246) | (0.257) | (0.576) | (0.273) | (0.252) | (0.128) |
| Weekends*Pandemic declaration*Risk-taking | 4.02*** | 3.46*** | 5.03*** | 3.97*** | 3.61*** | -0.99** |
| | (0.680) | (0.648) | (1.467) | (0.836) | (0.669) | (0.309) |
| Controls | Yes | Yes | Yes | Yes | Yes | Yes |
| Observations | 64800 | 64613 | 58858 | 62958 | 67073 | 58284 |
| Number of clusters | 796 | 785 | 738 | 761 | 798 | 741 |
| Prob. $> \chi^2$ | 0.000 | 0.000 | 0.000 | 0.000 | 0.000 | 0.000 |
| $R^2$-between | 0.850 | 0.576 | 0.389 | 0.810 | 0.775 | 0.809 |
| $R^2$-within | 0.633 | 0.470 | 0.283 | 0.272 | 0.529 | 0.631 |
| $R^2$-overall | 0.813 | 0.542 | 0.345 | 0.698 | 0.736 | 0.770 |

Notes: **Robust 1** = regions with at least one censored values on the outcome mobility measures excluded. **Robust 2** = regions with at least one censored values on any mobility measures excluded. **Robust 3** = government response indicators recoded as no measures taken if policy is not applied countrywide. Random-effects GLS regression estimates. Standard errors (clustered at regional level) in parentheses. † $p < .10$; * $p < .05$; ** $p < .01$; *** $p < .001$. We controlled for the day since first confirmed death, share of population over 65, number of confirmed cases (in logs), population density, and the set of government response indicators in each regression. Reference categories are: *Before WHO declares COVID-19 as pandemic*, *Weekdays* and *No measures taken*.

**Table S11 | Robustness checks on risk preference and share of population at risk interaction effect on mobility.**

| **Robust 1** | Retail & recreation | Grocery & pharmacy | Parks | Transit stations | Workplaces | Residential |
|---|---|---|---|---|---|---|
| Risk-taking | 6.46* | -1.77 | 17.2*** | 3.19 | 1.74 | 1.84** |
|  | (2.593) | (2.191) | (4.404) | (3.354) | (1.891) | (0.712) |
| % Population 65+ | -0.40* | -0.86*** | -1.97*** | -0.39* | -0.22* | 0.100† |
|  | (0.176) | (0.164) | (0.371) | (0.156) | (0.0999) | (0.0547) |
| Risk-taking* % Population 65+ | -0.37* | 0.13 | -0.83* | -0.17 | -0.090 | -0.17** |
|  | (0.185) | (0.167) | (0.363) | (0.233) | (0.123) | (0.0597) |
| Controls | Yes | Yes | Yes | Yes | Yes | Yes |
| Observations | 58340 | 57571 | 51715 | 57764 | 63041 | 44546 |
| Number of clusters | 688 | 678 | 609 | 680 | 744 | 524 |
| Prob. $> \chi^2$ | 0.000 | 0.000 | 0.000 | 0.000 | 0.000 | 0.000 |
| $R^2$-between | 0.861 | 0.600 | 0.403 | 0.827 | 0.788 | 0.806 |
| $R^2$-within | 0.417 | 0.431 | 0.348 | 0.169 | 0.513 | 0.496 |
| $R^2$-overall | 0.799 | 0.558 | 0.374 | 0.704 | 0.750 | 0.762 |
| **Robust 2** | Retail & recreation | Grocery & pharmacy | Parks | Transit stations | Workplaces | Residential |
| Risk-taking | 8.32*** | -1.82 | 14.9** | 4.50* | -0.95 | 2.24** |
|  | (1.922) | (1.737) | (4.687) | (2.041) | (1.279) | (0.745) |
| % Population 65+ | -0.78*** | -0.85*** | -2.13*** | -0.40* | -0.18 | 0.10† |
|  | (0.200) | (0.174) | (0.416) | (0.185) | (0.123) | (0.0570) |
| Risk-taking* % Population 65+ | -0.55** | 0.15 | -0.52 | -0.25 | 0.067 | -0.19** |
|  | (0.176) | (0.149) | (0.416) | (0.177) | (0.106) | (0.0630) |
| Controls | Yes | Yes | Yes | Yes | Yes | Yes |
| Observations | 41146 | 41146 | 41146 | 41146 | 41146 | 41146 |
| Number of clusters | 484 | 484 | 484 | 484 | 484 | 484 |
| Prob. $> \chi^2$ | 0.000 | 0.000 | 0.000 | 0.000 | 0.000 | 0.000 |
| $R^2$-between | 0.849 | 0.571 | 0.391 | 0.840 | 0.771 | 0.807 |
| $R^2$-within | 0.589 | 0.572 | 0.376 | 0.325 | 0.639 | 0.518 |
| $R^2$-overall | 0.804 | 0.571 | 0.378 | 0.773 | 0.752 | 0.765 |
| **Robust 3** | Retail & recreation | Grocery & pharmacy | Parks | Transit stations | Workplaces | Residential |
| Risk-taking | 4.23 | -5.50* | 6.36 | -1.40 | -1.28 | 2.53*** |
|  | (2.691) | (2.358) | (4.232) | (3.551) | (1.940) | (0.661) |
| % Population 65+ | 0.27† | -0.31† | -0.016 | 0.24 | 0.080 | -0.30*** |
|  | (0.147) | (0.175) | (0.473) | (0.175) | (0.0992) | (0.0528) |
| Risk-taking* % Population 65+ | -0.046 | 0.48** | -0.026 | 0.27 | 0.17 | -0.27*** |
|  | (0.163) | (0.159) | (0.345) | (0.209) | (0.114) | (0.0446) |
| Controls | Yes | Yes | Yes | Yes | Yes | Yes |
| Observations | 64800 | 64613 | 58858 | 62958 | 67073 | 58284 |
| Number of clusters | 796 | 785 | 738 | 761 | 798 | 741 |
| Prob. $> \chi^2$ | 0.000 | 0.000 | 0.000 | 0.000 | 0.000 | 0.000 |
| $R^2$-between | 0.846 | 0.568 | 0.386 | 0.808 | 0.768 | 0.802 |
| $R^2$-within | 0.633 | 0.471 | 0.284 | 0.271 | 0.527 | 0.640 |
| $R^2$-overall | 0.809 | 0.539 | 0.344 | 0.697 | 0.730 | 0.768 |

Notes: **Robust 1** = regions with at least one censored values on the outcome mobility measures excluded. **Robust 2** = regions with at least one censored values on any mobility measures excluded. **Robust 3** = government response indicators recoded as no measures taken if policy is not applied countrywide. Random-effects GLS regression estimates. Standard errors (clustered at regional level) in parentheses. † $p < .10$; * $p < .05$; ** $p < .01$; *** $p < .001$. We controlled for weekend dummy, pandemic declaration dummy, days since first confirmed death, number of confirmed cases (in logs), population density, and the set of government response indicators in each regression. Reference categories are: *Before WHO declares COVID-19 as pandemic*, *Weekdays* and *No measures taken*.